\documentclass[usenatbib]{mnras}
\hypersetup{draft}

\usepackage{graphicx}
\usepackage{savesym}
\savesymbol{tablenum}
\usepackage{siunitx}

\usepackage{amsmath}
\usepackage{amssymb}
\PassOptionsToPackage{hyphens}{url}\usepackage{hyperref}
\usepackage{breakurl}
\usepackage[xindy, toc, hyperfirst=false, nolist, nostyles]{glossaries}

\glsdisablehyper
\usepackage{comment}

\usepackage{booktabs}
\usepackage{longtable}

\usepackage{xspace}

\newcommand{\msun}{\hbox{$M_{\odot}$}}

\newglossaryentry{vrad}{name={radial velocity~}, text={radial velocity}, symbol={\ensuremath{v_\textrm{rad}}}, description={radial velocity}, sort=vrad}
\newglossaryentry{vrot}{name={stellar rotation~}, name={stellar rotation}, symbol={\ensuremath{v_\textrm{rot}}}, description={radial velocity}, sort=vrot}

\newcommand{\ra}[4]{\hbox{\ensuremath{\alpha=#1^h#2^m#3\overset{s}{.}#4}}}
\newcommand{\de}[3]{\hbox{\ensuremath{\delta=#1^{\circ}#2\arcmin#3\arcsec}}}

\newcommand{\xray}{X-ray}

\newcommand{\Ni}[1][56]{\ensuremath{^{#1}\textrm{Ni}}}
\newcommand{\Co}[1][56]{\ensuremath{^{#1}\textrm{Co}}}

\newcommand{\nucl}[2]{\ensuremath{^{#2}\textrm{#1}}}

\newglossaryentry{angstrom}{name=\AA, description={unit of length $10^{-10}$\,m}, sort=angstrom}
\newglossaryentry{nir}{name=NIR,description={near infrared},first = {near infrared (NIR)}}
\newglossaryentry{psf}{name=PSF,description={point-spread function},first = {point-spread function (PSF)}}
\newglossaryentry{fwhm}{name=FWHM,description={Full Width Half Maximum},first = {FWHM}}
\newglossaryentry{rms}{name=RMS,description={Root Mean Square},first = {RMS}}
\newglossaryentry{signalnoise}{name=S/N,description={signal to noise}}
\newglossaryentry{uv}{name=UV,description={ultra violet},first = {ultra violet (UV)}}
\newglossaryentry{halpha}{name=\ensuremath{\textrm{H}\alpha}, description={First line of the Balmer series at 6563\,\AA}, sort=halpha}
\newglossaryentry{mgb}{name={Mg \textsc{i} b}, description={Triplet at 5167\,\AA, 5173\,\AA and 5184\,\AA}}
\newglossaryentry{sobolevapprox}{name={Sobolev approximation}, description={Lines are approximation with an infinitley thin interaction region \citep[e.g. no broadening][]{1960mes..book.....S}}, first={Sobolev approximation }}
\newglossaryentry{radeq}{name={radiative equilibrium}, description={The net flux of energy between matter and radiation field is zero}}
\newglossaryentry{nebularapprox}{name={nebular approximation}, description={Assumes that the plasma condition are controlled by a central radiation source. The radiation field decreases with the distance to the source by geometrical dilution. See \citet{1978stat.book.....M} for details}}
\newglossaryentry{modnebularapprox}{name={modified nebular approximation}, description={In contrast to \gls{nebularapprox} where only geometrical dilution is taken into account, the modified nebular approximation also takes dilution by other radiative processes into account }, first={modified nebular approximation}, parent=nebularapprox}
\newglossaryentry{thompsonscat}{name={Thomson scattering}, description={Scattering of photons on low energy electrons}}
\newglossaryentry{lte}{name={LTE}, description={Local Thermodynamic Equilibrium}, first={local thermodynamic equilibrium (LTE)}}
\newglossaryentry{lsr}{name={LSR}, description={Local Standard of Rest}, first={\textit{local standard of rest} (LSR)}}
\newglossaryentry{mc}{name={MC}, description={Monte Carlo}, first={\textit{Monte Carlo} (MC)}}
\newglossaryentry{wcs}{name={WCS}, description={world coordinate system}, first={world coordinate system (WCS)}}
\newglossaryentry{cmf}{name=CMF, text=CMF, first=Comoving Frame (CMF henceforth), description={Comoving Frame}}

\newglossaryentry{uvoir}{name=UVOIR, text=UVOIR, first=UV/optical/Near-IR (UVOIR), description={UV/optical/Near-IR}}

\newglossaryentry{sfit}{name=SFIT, text=\textsc{sfit}, description={spectral fitting program for hot stars \citep{2001A&A...376..497J}}, first={\textsc{sfit} \citep{2001A&A...376..497J}}}
\newglossaryentry{iraf}{name=IRAF, text=\textsc{iraf}, description={Image Reduction and Analysis Facility maintained by NOAO}, first={\textsc{iraf}\protect\footnote{IRAF: the Image Reduction and Analysis Facility is distributed by the National Optical Astronomy Observatory, which is operated by the Association of Universities for Research in Astronomy (AURA) under cooperative agreement with the National Science Foundation (NSF).}}}
\newglossaryentry{pyraf}{name=PyRAF, text=\textsc{PyRAF}, description={Python wrap of \gls{iraf} maintained by STSCI}, first=\textsc{PyRAF} \protect\footnote{PyRAF is a product of the Space Telescope Science Institute, which is operated by AURA for NASA.}}
\newglossaryentry{astropy}{name=ASTROPY, text=\textsc{astropy}, description=\textsc{astropy} framework, first = \textsc{astropy} \citep{2013A&A...558A..33A}}
\newglossaryentry{numpy}{name=NUMPY, text=\textsc{numpy}, description=\textsc{numpy} framework, first = \textsc{numpy} \citep{walt2011numpy}}
\newglossaryentry{scipy}{name=SCIPY, text=\textsc{scipy}, description=\textsc{scipy} framework, first = \textsc{scipy} \citep{Jones:2001fk}}
\newglossaryentry{matplotlib}{name=matplotlib, text=\textsc{matplotlib}, description=\textsc{matplotlib} framework, first = \textsc{matplotlib} \citep{hunter2007matplotlib}}
\newglossaryentry{pandas}{name=pandas, text=\textsc{pandas}, description=\textsc{pandas} framework, first = \textsc{pandas} \citep{mckinney2010data}}
\newglossaryentry{ipython}{name=ipython, text=\textsc{ipython}, description=\textsc{ipython} framework, first = \textsc{ipython} \citep{perez2007ipython}}
\newglossaryentry{jupyter}{name=jupyter, text=\textsc{jupyter}, description=\textsc{jupyter} framework, first = \textsc{jupyter} \citep{kluyver2016jupyter,perez2015project,ragan2014jupyter}}
\newglossaryentry{aplpy}{name=aplpy, text=\textsc{aplpy}, description=\textsc{aplpy} framework, first = \textsc{aplpy} \citep{2012ascl.soft08017R}}

\newglossaryentry{moog}{name=MOOG,text={\textsc{moog}}, description={spectral synthesis software \citep{1973ApJ...184..839S}}, first={\textsc{Moog} \citep{1973ApJ...184..839S}}}
\newglossaryentry{atlas9}{name=ATLAS9,description={grid of stellar atmospheres \citep{2004astro.ph..5087C}}, first={ATLAS9 \citep{2004astro.ph..5087C}}}
\newglossaryentry{vald}{name=VALD,description={Vienna Atomic Line Database \citep{2000BaltA...9..590K}}, first={Vienna Atomic Line Database \citep[VALD;][]{2000BaltA...9..590K}}}
\newglossaryentry{sextractor}{name=SExtractor, text=\textsc{SExtractor}, description={Source Extractor photometry program \citep{1996A&AS..117..393B}}, first={\textsc{SExtractor} \citep{1996A&AS..117..393B}}}

\newglossaryentry{swarp}{name=SWarp, text=\textsc{SWarp}, description={SWarp \citep{2002ASPC..281..228B}}, first={\textsc{SWarp} \citep{2002ASPC..281..228B}}}

\newglossaryentry{astrodrizzle}{name=AstroDrizzle, text=\textsc{AstroDrizzle}, description={AstroDrizzle \citep{2012drzp.book.....G}}, first={\textsc{AstroDrizzle} \citep{2012drzp.book.....G}}}

\newglossaryentry{idl}{name=IDL,text={\textsc{idl}}, description={Interactive Data Language}}
\newglossaryentry{makee}{name=MAKEE,text=\textsc{makee}, description={MAuna Kea Echelle Extraction by Tom Barlow available}}
\newglossaryentry{minuit}{name=MINUIT,text={\textsc{minuit}}, description={collection of numerical optimization tools \citep{James:1975dr}}}
\newglossaryentry{migrad}{name=MIGRAD,text={\textsc{migrad}}, description={numerical gradient optimization tools - part of \gls{minuit}}}
\newglossaryentry{dolphot}{name=DOLPHOT, text=\textsc{dolphot}, description=photometry package for HST, first=\textsc{dolphot} \citep{2000PASP..112.1383D}}
\newglossaryentry{synphot}{name=synphot, text={\textsc{synphot}}, description={synthetic photometry package from STSCI}, first={\textsc{synphot}\protect\footnote{\textsc{synphot} is a product of the Space Telescope Science Institute, which is operated by AURA for NASA.}}}
\newglossaryentry{chianti}{name=CHIANTI, text=CHIANTI, description= CHIANTI Database 7.1, first =CHIANTI 7.1 \citep{1997A&AS..125..149D,2012ApJ...744...99L}}
\newglossaryentry{synpp}{name=SYNPP, text=SYN++, description= SYN++ software, first =SYN++ \citep{2011PASP..123..237T}}
\newglossaryentry{tardis}{name=TARDIS, text=\textsc{tardis}, description= TARDIS MC code, first = {\textsc{tardis} \citep{2014MNRAS.440..387K}}}

\newglossaryentry{artis}{name=ARTIS, text=\textsc{artis}, description= ARTIS MC code, first = \textsc{artis} \citep{2009MNRAS.398.1809K}}
\newglossaryentry{sedona}{name=SEDONA, text=\textsc{sedona}, description= Sedona MC code, first = \textsc{sedona} \citep{2006ApJ...651..366K}}
\newglossaryentry{mlmc}{name=MLMC, text=ML93, description= Mazzali Lucy Monte Carlo, first ={Mazzali \& Lucy (1993, ML93) code}}
\newglossaryentry{starkit}{name=STARKIT, text=\textsc{starkit}, description= TARDIS MC code, first = {\textsc{starkit} \citep{wolfgang_kerzendorf_2015_28016}}}

\newglossaryentry{pyne}{name=PYNE, text=\textsc{pyne}, description= PYNE code, first = {\textsc{pyne} \citep{Scopatz2012a}}}
\newglossaryentry{multinest}{name=MULTINEST, text=\textsc{MultiNest}, description=MultiNest, first={\textsc{MultiNest} \citep{2009MNRAS.398.1601F}}}

\newglossaryentry{ads}{name=ADS ,description=ADS, first={NASA Astrophysics Data System \citep{2000A&AS..143...41K}}}
\newglossaryentry{2mass}{name=2MASS,description={Two Micron All Sky Survey \citep{2006AJ....131.1163S}}, first={Two Micron All Sky Survey \citep{2006AJ....131.1163S}}}
\newglossaryentry{nomad}{name=NOMAD,first={Naval Observatory Merged Astrometric Dataset \citep[NOMAD; ][]{2005yCat.1297....0Z}}, description={Naval Observatory Merged Astrometric Dataset}}
\newglossaryentry{wifes}{name=WIFES, text=\textsc{WiFeS}, first={\textsc{WiFeS} \citep{2007Ap&SS.310..255D}},  description={Wide Field Spectrograph - \gls{ifu} mounted on the 2.3\,m telescope at Siding Spring Observatory}}
\newglossaryentry{scp}{name=SCP,description={Supernova Cosmology Project, led by Saul Perlmutter}, first={Supernova Cosmology Project (SCP)}}
\newglossaryentry{hzsns}{name=HZSNS,description={High Z Supernova Search, led by Brian Schmidt}, first={High Z Supernova Search (HZSNS)}}
\newglossaryentry{vlt}{name=VLT,description={Very Large Telescope located on Cerro Paranal (Chile)}, first={Very Large Telescope (VLT)}}
\newglossaryentry{flames}{name=FLAMES,description={Multi-object, intermediate and high resolution spectrograph mounted on the  \gls{vlt}}}
\newglossaryentry{hires}{name=HIRES, description={High Resolution Echelle Spectrometer mounted on the Keck Telescope}, first={High Resolution Echelle Spectrometer \citep[HIRES;][]{1994SPIE.2198..362V}}}
\newglossaryentry{lris}{name=LRIS,description={Low Resolution Imaging Spectrometer mounted on the Keck Telescope}, first={Low-Resolution Imaging Spectrometer \citep[LRIS;][]{Oke95}}}

\newglossaryentry{essence}{name=ESSENCE,description={The `Equation of State: SupErNovae trace Cosmic Expansion' project \citep[ESSENCE;][]{2002AAS...201.7809G}}, first={`The Equation of State: SupErNovae trace Cosmic Expansion' \citep[ESSENCE;][]{2002AAS...201.7809G}}}
\newglossaryentry{ifu}{name=IFU,description={Optical instrument combining spectrographic and imaging capabilities, used to obtain spatially resolved spectra}, first={Integral Field Unit (IFU)}, firstplural={Integral Field Units (IFUs)}}

\newglossaryentry{besancon}{name=Besan\c{c}on Model, description={Model of stellar population synthesis of the Galaxy, including kinematics.}}

\newglossaryentry{int}{name=INT,description={Isaac Newton 2.5\,m Telescope}, first={Isaac Newton 2.5\,m Telescope (INT)}}
\newglossaryentry{iau}{name=IAU,description={International Astronomical Union}, first={IAU}}
\newglossaryentry{chandra}{name=Chandra,description={Chandra \xray\ Observatory (space-based)}}
\newglossaryentry{hst}{name=HST,description={Hubble Space Telescope}}
\newglossaryentry{hst.wfpc2}{name=WFPC2,description={Wide-Field Planetary Camera 2 mounted on the \gls{hst}}, first={Wide-Field Planetary Camera 2 (WFPC2)}}
\newglossaryentry{hst.acs}{name=ACS,description={Advanced Camera for Surveys mounted on the \gls{hst}}, first={Advanced Camera for Surveys (ACS)}}
\newglossaryentry{hst.wfc3}{name=WFC3,description={Wide-Field Camera 3 mounted on the \gls{hst}}, first={Wide-Field Camera 3 (WFC3)}}
\newglossaryentry{hst.cte}{name=CTE, description={charge transfer efficiency (CTE)}, first={charge transfer efficiency \citep[CTE; see ][for a description]{2009acs..rept....1C}}}

\newglossaryentry{snls}{name=SNLS,description={Supernova Legacy Survey \citep{2003AAS...203.8209P}}, first={Supernova Legacy Survey \citep[SNLS;][]{2003AAS...203.8209P}}}
\newglossaryentry{dass}{name=DASS, description={Digitized Astronomy Supernova Survey \citep{1975PASP...87..565C}}, first={Digitized Astronomy Supernova Survey \citep[DASS;][]{1975PASP...87..565C}}}
\newglossaryentry{bait}{name=BAIT, description={Berkley Automatic Imaging Telescope \citep{1993PASP..105.1164R}}, first={Berkley Automatic Imaging Telescope \citep[BAIT;][]{1993PASP..105.1164R}}}
\newglossaryentry{kait}{name=KAIT, description={Katzman Automatic Imaging Telescope \citep{2001ASPC..246..121F}}, first={Katzman Automatic Imaging Telescope \citep[KAIT;][]{2001ASPC..246..121F}}}
\newglossaryentry{loss}{name=LOSS, description={Lick Observatory Supernova Search  \citep{2000AIPC..522..103L}}, first={Lick Observatory Supernova Search \citep[LOSS;][]{2000AIPC..522..103L}}}
\newglossaryentry{ctss}{name=CTSS,description={Cal\'{a}n/Tololo Supernova Survey \citep{1993AJ....106.2392H}}, first={Cal\'{a}n/Tololo supernova survey \citep[CTSS;][]{1993AJ....106.2392H}}}
\newglossaryentry{ctio}{name= CTIO, description={Cerro Tololo Inter-American Observatory}, first={Cerro Tololo Inter-American Observatory (CTIO)}}
\newglossaryentry{ptf}{name=PTF, description={Palomar Transient Factory \citep{2009PASP..121.1334R}}, first={Palomar Transient Factory \citep[PTF;][]{2009PASP..121.1334R}}}
\newglossaryentry{batse}{name=BATSE, description={Burst and Transient Source Experiment mounted on the Compton Gamma Ray Observatory}, first={Burst and Transient Source Experiment (BATSE)}}
\newglossaryentry{bepposax}{name=BeppoSAX, description={\xray\ satellite named in honor of Giuseppe "Beppo" Occhialini}}
\newglossaryentry{rosat}{name=ROSAT, description={short for R\"{o}ntgensatellit}, first={ROSAT}}
\newglossaryentry{hete2}{name=HETE2, description={High Energy Transient Explorer}, first={High Energy Transient Explorer (HETE)}}
\newglossaryentry{ska}{name=SKA, description={Square Kilometre Array}, first={Square Kilometre Array (SKA)}}

\newglossaryentry{gnirs}{name=GNIRS, description={Gemini Near InfraRed Spectrograph mounted on the Gemini North Telescope}}
\newglossaryentry{gmosn}{name=GMOS, description={Gemini Multi Object Spectrograph mounted on the
 Gemini North Telescope}, first={GMOS \citep[Gemini Multi Object Spectrograph;][]{2004PASP..116..425H}}}
\newglossaryentry{swift}{name=Swift, description={Swift Gamma-Ray Burst Mission}}
\newglossaryentry{vla}{name=VLA, description={Very Large Array radio telescope located in North America}, first={Very Large Array (VLA)}}
\newglossaryentry{evla}{name=EVLA, description={Extended Very Large Array radio telescope located in North America}, first={Extended Very Large Array (EVLA)}}
\newglossaryentry{sdss}{name=SDSS, description={Sloan Digital Sky Survey}}
\newglossaryentry{dss}{name=DSS, description={Digitized Sky Survey}}
\newglossaryentry{skymapper}{name=SkyMapper, description={SkyMapper telescope \citep{2007PASA...24....1K}}, first={SkyMapper \citep{2007PASA...24....1K}}}
\newglossaryentry{panstarrs}{name=PanSTARRS, description={Panoramic Survey Telescope \& Rapid Response System \citep{2004SPIE.5489...11K}}, first={Panoramic Survey Telescope \& Rapid Response System \citep[PanSTARRS;][]{2004SPIE.5489...11K}}}
\newglossaryentry{lsst}{name=LSST, description={Large Synoptic Survey Telescope}, first={Large Synoptic Survey Telescope \citep[LSST;][]{2006AAS...209.8604P}}}
\newglossaryentry{ppmxl}{name=PPMXL, description={PPMXL Catalog of Positions and Proper Motions on the ICRS \citep{2010AJ....139.2440R}}}
\newglossaryentry{gaia}{name=GAIA, description={Global Astrometric Interferometer for Astrophysics \citep{2001A&A...369..339P}}, first={Global Astrometric Interferometer for Astrophysics \citep[GAIA;][]{2001A&A...369..339P}}}
\newglossaryentry{ligo}{name=LIGO, description={Laser Interferometer Gravitational Wave Observatory}, first={Laser Interferometer Gravitational Wave Observatory \citep[LIGO;][]{1992Sci...256..325A}}}
\newglossaryentry{aligo}{name=Advanced LIGO, description={Advanced LIGO}, sort=ligo2}
\newglossaryentry{lisa}{name=LISA, description={Laser Interferometer Space Antenna \citep{1994ESAJ...18..219J}}, first={Laser Interferometer Space Antenna \citep[LISA;][]{1994ESAJ...18..219J}}}
\newglossaryentry{ccd}{name=CCD,description={Charged Coupled Device}, first={charged coupled device (CCD)}, firstplural={charged coupled devices (CCDs)}}

\newcommand{\sn}[2]{SN~#1#2\xspace}

\newglossaryentry{irc}{name=IRC, text={IRC}, description={infrared catastrophe}, first={infrared catastrophe \citep[IRC;][]{1980PhDT.........1A}}}

\newglossaryentry{sn}{name=Supernova, text={SN}, plural={SNe}, description={exploding star}, nonumberlist=true, first={supernova (SN)}, firstplural={supernovae (SNe)}}
\newglossaryentry{snia}{name=Type~Ia (SN~Ia), text={SN~Ia}, description={Thermonuclear explosion of a white dwarf - spectra show no hydrogen but a strong silicon line},first={Type~Ia supernova (SN~Ia)}, firstplural={Type Ia supernovae (SNe~Ia)}, plural={SNe~Ia}, parent=sn, nonumberlist=true}
\newcommand{\sneia}{\glspl*{snia}\xspace}
\newcommand{\snia}{\gls*{snia}\xspace}

\newglossaryentry{branchnormal}{name={branch-normal}, text=\textit{Branch-normal}, description={Large homogeneous class of Type Ia Supernovae, defined in \citet{1993AJ....106.2383B}}, first={\textit{Branch-normal} SNe Ia \citep{1993AJ....106.2383B}}, parent=snia} 
\newglossaryentry{91t}{name={91T-like}, description={Luminous class of Type Ia supernovae similar to \sn{1991}{T} \citep{1992AJ....103.1632P}} , first={91T-like}, parent=snia} 
\newglossaryentry{91bg}{name={91bg-like}, description={Faint class of Type Ia supernovae similar to \sn{1991}{bg} \citep{1992AJ....104.1543F}}, first={91bg-like}, parent=snia} 
\newglossaryentry{02cx}{name={02cx-like}, description={Peculiar class of Type Ia supernovae similar to \sn{2002}{cx} \citep{2003PASP..115..453L}}, first={02cx-like \sneia\ \citep{2003PASP..115..453L}}, parent=snia} 

\newglossaryentry{snibc}{name=Type~Ib/c, text={SN~Ib/c}, description={Collapse of the core of a massive star -  spectrum shows no hydrogen and no silicon line},first={Type~Ib/c supernova (SN~Ib/c)}, firstplural={Type~Ib/c supernovae (SNe~Ib/c)}, plural={SNe~Ib/c}, parent=sn}

\newglossaryentry{snib}{name=Type~Ib, text={SN~Ib}, description={Spectrum shows no hydrogen and no silicon, but helium line},first={Type Ib supernova (SN~Ib)}, firstplural={Type~Ib supernovae (SNe~Ib)}, plural={SNe~Ib}, parent=snibc}

\newglossaryentry{snic}{name=Type~Ic, text={SN~Ic}, description={Spectrum shows no hydrogen, no silicon and no helium line},first={Type~Ic supernova (SN~Ic)}, firstplural={Type~Ic supernovae (SNe~Ic)}, plural={SNe~Ic}, parent=snibc}

\newglossaryentry{snii}{name=Type~II, text={SN~II}, description={Collapse of the core of a massive star - spectrum shows strong hydrogen line},first={Type~II supernova (SN~II)}, firstplural={Type~II supernovae (SNe~II)}, plural={SNe~II}, parent=sn}

\newglossaryentry{sniib}{name=Type~IIb, text={SN~IIb}, description={Spectrum shows hydrogen and helium lines},first={Type~IIb supernova (SN~IIb)}, firstplural={Type~IIb supernovae (SNe~IIb)}, plural={SNe~IIb}, parent=snii}

\newglossaryentry{sniip}{name=Type~II~Plateau (Type IIP), text={SN~IIP}, description={Lightcurve shows plateau},first={Type~IIP supernova (SN~IIP)}, firstplural={Type~II Plateau supernovae \citep[SNe~IIP;][]{1979A&A....72..287B}}, plural={SNe~IIP}, parent=snii}

\newglossaryentry{sniil}{name=SN~II~Linear, text={SN~IIL}, description={Lightcurve shows no plateau, but linear decline},first={Type~IIL supernova (SN~IIL)}, firstplural={Type~II~Linear supernovae \citep[SNe~IIL;][]{1990MNRAS.244..269S}}, plural={SNe~IIL}, parent=snii}

\newglossaryentry{sniin}{name=Type II narrow-lined (Type IIn), description={Spectrum shows narrow lines},first={Type~II~narrow-lined supernova (SN IIn)}, firstplural={Type~IIn supernovae (SNe~IIn)}, plural={SNe~IIn}, parent=snii}

\newglossaryentry{snr}{name=Remnant (SNR), text=SNR, description={Remnant left visible post-explosion}, first={supernova remnant (SNR)}, firstplural={supernova remnants (SNRs)}, parent=sn}

\newglossaryentry{dtd}{name=DTD,description={delay time distribution - expected supernova rate over time after a brief outburst of starformation},first={delay time distribution (DTD)}, firstplural={delay time distributions (DTDs)}, plural=DTDs}

\newglossaryentry{hvg}{name=HVG,description={high velocity gradient - Type Ia supernovae with a fast evolution of photospheric velocity},first={high velocity group (HVG)}, firstplural={high velocity groups (HVGs)}, plural=HVGs, parent=snia}

\newglossaryentry{lvg}{name=LVG,description={low velocity gradient - Type Ia supernovae with a slow evolution of photospheric velocity},first={low velocity group (LVG)}, firstplural={low velocity groups (LVGs)}, plural=LVGs, parent=snia}

\newglossaryentry{wd}{name=white dwarf (WD), text=WD, description={White Dwarf - extremely dense stellar remnant}, first={white dwarf (WD)}}
\newglossaryentry{onemgwd}{name= Oxygen/Neon (ONe), text={ONe-WD},description={Oxygen/Neon White Dwarf}, first={oxygen/neon White Dwarf (ONe-WD)}, parent=wd}
\newglossaryentry{cowd}{name=carbon/oxygen (CO), text={CO-WD}, description={carbon/oxygen white dwarf}, first={carbon/oxygen white dwarf (CO-WD)}, firstplural = {carbon/oxygen white dwarfs (CO-WDs)}, parent=wd}

\newglossaryentry{sds}{name=SD-Scenario,description={single-degenerate scenario (single white dwarf accreting from non-degenerate companion)}, first={single-degenerate scenario (SD-scenario)}}

\newglossaryentry{dds}{name=DD-Scenario, description={double degenerate scenario (merging of two white dwarfs)}, first={double-degenerate scenario (DD-scenario)}}

\newglossaryentry{sss}{name=SSS, text={supersoft \xray\ source}, description={supersoft \xray\ source - believed to be emitted by nuclear fusion on a white dwarf's surface}}

\newglossaryentry{amcvn}{name=AM CVn, description={AM Canum Venaticorum star (white dwarf accreting hydrogen poor matter from a companion star; see \cite{2005ASPC..330...27N})}}

\newglossaryentry{rlof}{name=RLOF, description={Roche Lobe Overflow (see \citet{1971ARA&A...9..183P} for a more detailed description)}, first={Roche-lobe overflow (RLOF)}}

\newglossaryentry{mchan}{name={Chandrasekhar mass~}, text={Chandrasekhar~mass}, symbol={\ensuremath{M_\textrm{Chan}}}, plural={Chandrasekhar~masses}, description={Mass when the core of a star collapses due to insufficient degeneracy pressure - for a white dwarf $\approx1.38\,M_\odot$ see \citet{1931ApJ....74...81C}}, first={Chandrasekhar~mass \citep[$M_\textrm{Chan}=1.38\,M_\odot$;][]{1931ApJ....74...81C}}, sort=mchan}

\newglossaryentry{w7}{name={W7 model},description={W7 model \citep{1984ApJ...286..644N}},first = {W7 model \citep{1984ApJ...286..644N}}}

\newglossaryentry{ew}{name=Equivalent Width, text={EW}, description={width of a rectangle that has the same area as a spectral line when taken to zero flux}, first={equivalent width (EW)}, firstplural={equivalent widths (EWs)}}
\newglossaryentry{agb}{name=AGB,description={Asymptotic Giant Branch}, first={Asymptotic Giant Branch (AGB)}}
\newglossaryentry{cmb}{name=CMB,description={Cosmic Microwave Background}}
\newglossaryentry{csm}{name=CSM,description={Circumstellar Medium}, first={circumstellar medium (CSM)}}
\newglossaryentry{csi}{name=CSI,description={Circumstellar Interaction}, first={circumstellar interaction (CSI)}}
\newglossaryentry{ism}{name=ISM,description={Interstellar Medium}, first={interstellar medium (ISM)}}
\newglossaryentry{ige}{name=IGE,description={Iron Group Element}, first={iron group element (IGE)}, firstplural={iron group elements (IGEs)}}
\newglossaryentry{epm}{name=EPM,description={Expanding Photosphere Method \citep{1974ApJ...193...27K}}, first={Expanding Photosphere Method (EPM)}}
\newglossaryentry{aic}{name=AIC,description={Accretion Induced Collapse}, first={accretion induced collapse (AIC)}}
\newglossaryentry{ime}{name=IME,description={Intermediate Mass Element}, first={intermediate mass element (IME)}, firstplural={intermediate mass elements (IMEs)}}
\newglossaryentry{h0}{name=\ensuremath{H_0},description={Hubbles constant}}
\newglossaryentry{nse}{name=NSE,description={Nuclear Statistical Equilibrium}, first={nuclear statistical equilibrium (NSE)}}
\newglossaryentry{cdm}{name=CDM,description={Cold Dark Matter}}
\newglossaryentry{grb}{name=GRB,description={Gamma Ray Burst}, first={Gamma Ray Burst (GRB)}, firstplural={Gamma Ray Bursts (GRBs)}}
\newglossaryentry{donor}{name=donor,description={non-degenerate companion in the \gls{sds}}}
\newglossaryentry{mainsequence}{name=main sequence,description={main sequence star}}
\newglossaryentry{redgiant}{name=red giant,description={red giant star}}
\newglossaryentry{mlcs}{name=MLCS,description={Multicolor Light Curve Shape method \citep[MLCS;][]{1996ApJ...473...88R}}, first={Multicolor Light-Curve Shape method \citep[MLCS;][]{1996ApJ...473...88R}}}
\newglossaryentry{rsoph}{name=RS~Ophiuci ,description={white dwarf accreting from a red giant - assumed progenitor of the \gls{sds}}, sort=rsoph}
\newglossaryentry{usco}{name=U~Scorpii,description={white dwarf accreting from a main sequence star - assumed progenitor of the \gls{sds}}, sort=usco}
\newglossaryentry{rcw86}{name=RCW~86,description={supernova remnant sometimes associated with \sn{185}{}}, sort=rcw86}
\newglossaryentry{casa}{name=Cas~A,description={Cassiopeia A supernova remnant - probably a \gls{snib} event}}
\newglossaryentry{cepheid}{name=Cepheid,description={very luminous variable star with a strong luminosity period relationship}}
\newglossaryentry{urca}{name=Urca, text=\textit{Urca}, description={process predominatly contributing to cooling in stars. The \textit{Urca} process consists of alternating electron-capture and $\beta^{-}$ decay of two nuclei pairs.},sort=urca} 
\newglossaryentry{alphacen}{name=Alpha Centauri,description={one of the brightest stars in the night sky and a close binary}}
\newglossaryentry{pcygni}{name={P Cygni}, text={P Cygni},description={a hypergiant luminous blue variable with strong winds. Often referred to as a description for their line profiles showing a emission peak at the rest wavelength of the line and a blue-shifted absorption trough.}}

\newglossaryentry{teff}{name={effective temperature~}, text={effective temperature}, symbol={\ensuremath{T_\textrm{eff}}}, description={Temperature of a blackbody emitting the same total energy}, sort=teff}

\newglossaryentry{logg}{name={surface gravity~}, text={surface gravity}, symbol={\ensuremath{\textrm{log}\,g}}, description={gravity at the surface of a star}, sort=logg}
\newglossaryentry{feh}{name={metallicity~}, text={metallicity}, symbol=\textrm{[Fe/H]},description={iron abundance relative to the sun}, sort=feh}

\newglossaryentry{texp}{name={time since explosion~}, text={time since explosion}, text={time since explosion}, symbol={\ensuremath{t_{\rm exp}}},description={time since explosion (measured in days)}, sort=texp, first={time since explosion (\ensuremath{t_{\rm exp}})}}

\newglossaryentry{lmc}{name=LMC,description={Large Magellanic Cloud}, first={Large Magellanic Cloud (LMC)}, sort=lmc}
\newglossaryentry{smc}{name=SMC,description={Small Magellanic Cloud}, sort=smc}
\newglossaryentry{z}{name=\ensuremath{z},description={redshift}, sort=z}

\title[Extremely late photometry of SN~2011fe]{Extremely late photometry of the nearby SN~2011fe\footnote{Based on observations made with the NASA/ESA Hubble Space Telescope, obtained [from the Data Archive] at the Space Telescope Science Institute, which is operated by the Association of Universities for Research in Astronomy, Inc., under NASA contract NAS 5-26555.}}

\author[Wolfgang~E.~Kerzendorf et al.]{W.\,E.~Kerzendorf,$^{\! 1}$ 
C.~McCully,$^{\! 2,3}$
S.~Taubenberger,$^{\! 1}$
A.~Jerkstrand,$^{\! 4}$ 
I.~Seitenzahl,$^{\! 5,6,7}$ \newauthor
A.\,J.~Ruiter,$^{\! 6,7,5}$
J.~Spyromilio,$^{\! 1}$,
K.\,S.~Long$^{8,9}$ and
C.~Fransson$^{10}$\\
$^{1}$European Southern Observatory, Karl-Schwarzschild-Str.~2, 85748 Garching bei M\"{u}nchen, Germany\\
$^{2}$Department of Physics, University of California, Santa
Barbara, Broida Hall, Mail Code 9530, Santa Barbara, CA
93106-9530, USA\\
$^{3}$Las Cumbres Observatory, Global Telescope Network,
6740 Cortona Drive Suite 102, Goleta, CA 93117, USA\\
$^{4}$Max Planck Institute for Astrophysics, Garching bei M\"{u}nchen, Karl-Schwarzschild-Str.~1, Postfach 1317, D-85741 Garching, Germany\\
$^{5}$School of Physical, Environmental and Mathematical Sciences, University of New South Wales, Australian Defence Force Academy,\\
Canberra, ACT 2600, Australia\\
$^{6}$Research School of Astronomy and Astrophysics, Australian National University, Canberra, ACT 0200, Australia\\
$^{7}$ARC Centre of Excellence for All-sky Astrophysics (CAASTRO)\\
$^{8}$Space Telescope Science Institute, 3700 San Martin Drive, Baltimore, MD, 21218, USA\\
$^{9}$Eureka Scientific, Inc. 2452 Delmer Street, Suite 100, Oakland, CA 94602-3017\\
$^{10}$Department of Astronomy, The Oskar Klein Centre, Stockholm University, Alba Nova University Centre, SE-106 91 Stockholm, Sweden
}

\begin{document}


\maketitle

\begin{abstract}
Type~Ia supernovae are widely accepted to be the outcomes of thermonuclear explosions in white dwarf stars. However, many details of these explosions remain uncertain (e.g. the mass, ignition mechanism, and flame speed). Theory predicts that at very late times (beyond 1000\,d) it might be possible to distinguish between explosion models. Few very nearby supernovae can be observed that long after the explosion. The Type~Ia supernova SN\,2011fe located in M101 and along a line of sight with negligible extinction, provides us with the once-in-a-lifetime chance to obtain measurements that may distinguish between theoretical models. In this work, we present the analysis of photometric data of SN\,2011fe taken between 900 and 1600 days after explosion with Gemini and {\it HST}. At these extremely late epochs theory suggests that the light curve shape might be used to measure isotopic abundances which is a useful model discriminant. However, we show in this work that there are several currently not well constrained physical processes introducing large systematic uncertainties to the isotopic abundance measurement. We conclude that without further detailed knowledge of the physical processes at this late stage one cannot reliably exclude any models on the basis of this dataset. 
\end{abstract}


\section{Introduction}

\sneia\ have been very successful distance indicators partially due to the fact that they form a relatively homogeneous class. 

Specifically, this homogeneity is driven by the fact that the majority of the radiated energy is provided by the decay chain $\nucl{Ni}{56} \rightarrow
{\nucl{Co}{56}} \rightarrow
{\nucl{Fe}{56}}$ \citep{1969ApJ...157..623C}. The energy is released in $\gamma$-rays and positrons that interact through Compton scattering and deposit their energy in heating the free electron gas, ionisations and excitations. This isotope chain, the lack of hydrogen, and existence of intermediate mass elements point to explosive nucleosynthesis in highly degenerate matter present in massive CO white dwarfs (1\,\msun\ -- 1.4\,\msun). However, the precise mass range, ignition mechanism and flame propagation (deflagration, detonation, mixed scenarios) remain to be discovered.  

In this work, we will focus on two specific models: The traditional scenario that sees the self-ignition of a CO white dwarf at a mass of $\approx 1.38~\msun$ (known as Chandrasekhar mass models) and alternate scenarios that ignite mostly due to dynamical effects (such as mergers). \citet{2012ApJ...750L..19R} presents two representatives of these two classes: A model at 1.38~\msun\ (N100) and a violent merger of a 0.9~\msun\ and 1.1~\msun\ CO white dwarfs. Both scenarios result in similar observables due to the comparable amounts of \Ni\ produced and thus are not easily distinguishable.

Specifically, both the models are currently able to produce synthetic light curves and spectra that are consistent with early epoch observations of \sneia \citep[$\approx 30$ days post-explosion; e.g.][]{2012ApJ...750L..19R}. Thus early time observations can not be easily used to successfully discriminate between the competing scenarios. However, late time photometry (from a few hunded days onwards) may provide an avenue to determine the progenitor system. 

One of the main difference between the two models is the density at which a large fraction of the matter burns  ($>2 \times 10^8\ \textrm{g}\ \textrm{cm}^{-3}$ for the N100 model and $< 2 \times 10^8\ \textrm{g}\ \textrm{cm}^{-3}$ for the merger model). Both models produce similar amounts of \nucl{Ni}{56} but the \nucl{Ni}{57} and \nucl{Co}{55} yields are different for these two channels. This effect can be seen in very late light curves ${\gtrsim}1000$\,d when their daughter nuclei (\nucl{Co}{57} and \nucl{Fe}{55}) start to dominate the energy deposition \citep{2009MNRAS.400..531S,2011PrPNP..66..329S}. This difference leads to a slower decline of the light curve of the N100 model compared to the merger scenario at epochs ${\gtrsim}1000$\,d \citep{2012ApJ...750L..19R}. However, at these late times the supernova is roughly a million times dimmer than at peak and thus this experiment can only be performed on very close-by objects.

SN\,2011fe is one of the closest \sneia \citep[6.4~Mpc;][]{2011ApJ...733..124S} and is essentially unattenuated \citep[$A_V < 0.05$;][]{2013A&A...549A..62P}. It provides us with a unique opportunity to observe a spectroscopically `normal' \snia later than any other \snia \citep[for a detailed description of \sn{2011}{fe} see][]{2017ApJ...841...48S}, thus allowing us to test the predicted difference in light curve shapes at very late epochs. 

We note that \citet{2017ApJ...841...48S} and \cite{2017MNRAS.468.3798D} have used datasets that overlap with those used in this work but have drawn different conclusions on the progenitor system despite using similar methods. This is due to the different treatment of the systematic uncertainties that affect the late light curve modelling -- a particular focus of this work \citep[these systematics might affect similar studies on other supernovae; e.g. SN~2012cg and SN~2014J.][]{2016ApJ...819...31G, 2017arXiv170401431Y}.

The method is fundamentally simple: Assuming that the energy that is measured (in this case between 0.4 -- 1.6 micron) is proportional to the energy produced by the decay of the important isotopes \nucl{Ni}{56}, \nucl{Ni}{57}, and \nucl{Co}{55}, we can infer the presence and amount of these isotopes in the ejecta. The complication in the experiment is to determine the systematic effects that affect the proportionality of the amounts of the isotopes to the observed flux \citep[c.f.][]{2001ApJ...559.1019M}. 

There are several such effects that might lead to a deviation between \gls{uvoir} flux and the radioactively injected energy at very late times. \citet{2017ApJ...841...48S, 2017MNRAS.468.3798D} already discuss several possibilities (surviving companion, light echoes, CSM interactions) and suggest that none of these do contribute significant flux for this supernova. 

In this paper, we will focus on effects that are closely related to the decay radiation. Fast electrons and positrons are a crucial energy carriers at these late times. \citet{1993ApJ...405..614C} suggested that some of these energy carriers (positrons) may begin to escape from the SN without annihilating as early as a few hundred days after explosion, resulting in a departure from the exponential decline \citep[discussed in detail in][]{2014ApJ...796L..26K}. While some authors \citep[e.g.][]{1999ApJS..124..503M,2001ApJ...559.1019M} argue for moderate positron escape occurring in some \sneia, more recent publications that take essential near-IR corrections into account require almost complete trapping of positrons up to quite late epochs to explain observations \citep[e.g.][]{2004A&A...428..555S,2007A&A...470L...1S,2009A&A...505..265L}. This complete trapping can be explained by a tangled magnetic field, which in turn could be used to constrain the nature of the progenitor white dwarf. 

\citet[][]{1980PhDT.........1A} predicted that the optical and near-IR light curves will decline much more rapidly after ${\sim}700\,\mathrm{d}$, even if all positrons remain trapped due to an infrared catastrophe (IRC). The IRC is predicted to occur when the electron gas temperature drops below that required to collisionally excite optical and near-IR atomic transitions ($T {\lesssim}1500\,\mathrm{K}$), and cooling proceeds only via fine structure lines emitting in the far-IR. This cooling is constant while the energy input continues to decline exponentially. In \citet{1992MNRAS.255..671S}, the IRC was determined to have occured in the core-collapse SN\,1987A. However, such a dimming has never been observed for \sneia \citep[e.g.][]{2009A&A...505..265L} which remains a puzzle for late time light curves. 

\citet{2015ApJ...814L...2F} suggest that while a large fraction of the energy is radiated away by fine structure lines in the far-IR, the observed flux in the UVOIR bands comes from non-thermal excitation of Fe that produces UV photons, which in turn interact with the ejecta and produce the observed flux distribution. 

In this work, we expand the previous observational data set on SN\,2011fe and use a Bayesian framework to be able to incorporate prior knowledge quantitatively into the analysis process to give a consistent measurement of the isotopic ratios. 

In Section~\ref{sec:observations} we present our observations. In Section~\ref{sec:analysis} we discuss the analysis -- specifically the construction of the quasi-bolometric light curve and the light curve model. We discuss the various physical processes that lead to systematic uncertainties in Section~\ref{sec:discussion} and conclude the paper in Section~\ref{sec:conclusion}.

\section{Observations}
\label{sec:observations}
\subsection{HST Photometry}

The observations presented in this work are compiled from three different programs (PI: Kerzendorf -- GO 13824; PI: Shappee -- GO 14166, GO 13737) which imaged \sn{2011}{fe} (\ra{14}{03}{05}{7}, \de{+54}{16}{25.18}) between 1125 and 1623\,d after MJD$_\textrm{max}$ = 55814.51 \citep{2013A&A...554A..27P} using both \gls{hst.wfc3} and \gls{hst.acs}. In addition, we have included archival \gls{hst.acs} observations from GO 9490 (PI: Kuntz).

We generally use the ``FLC'' frames with \gls{hst.cte} correction throughout our work. For sub-frame images (our UVIS WFC3 images) could not be \gls{hst.cte} corrected. The systematics introduced by this are negligible as \sn{2011}{fe} is located very close to the amplifier. 

We did coarse registration of our single frames with our registered templates. For this step, we used \textsc{astroscrappy}\footnote{\url{https://github.com/astropy/astroscrappy} \citep[cf.][]{2001PASP..113.1420V}} to clean the raw frames to improve the matching process. We then selected the 120 brightest stars around \sn{2011}{fe} and matched those to stars in the reference image (Visit 4 - F438W drizzled frame). We then fit a TAN-projection WCS matrix (four parameters) transforming between the image coordinates of the stars within the unregistered frame and the WCS coordinates in the new frame using the differential evolution algorithm (implemented in \textsc{scipy.optimize.differential\_evolution}). The registered images produced from this technique had a median offset $\approx 0.1$~pixel for each of the matching stars.

We then did a fine registration using the \textsc{TWEAKREG} algorithm before we combined frames. The final cosmic ray rejection was done on raw FLC frames and thus combined the dithered observations using \gls{astrodrizzle} (combining frames that come from the same visit, instrument and filter). 

We ran \gls{dolphot} -- a PSF photometry package -- on all images: \gls{dolphot} runs on the individual frames, but registers to a template drizzled frame; photometry is performed on all frames simultaneously. For a source to be detected it must be detected at least in one frame with a 2.5-$\sigma$ threshold or on the combined with a 3.5-$\sigma$ threshold. If a source is detected it is added to the catalogue and PSF photometry is performed in this position on all frames. For this the frames must be astrometrically matched. This technique allows us to extract the most accurate photometry for the supernova even in images where nearby stars are close to it.

On the resulting catalogue we performed a standard cut of object type (1, 2 as described in the \gls{dolphot} manual), in sharpness between $-0.3$ and 0.3 as well as crowding $<0.5$.

We then produced the final photometry for each star by combining the individual measurements using standard uncertainty propagation.

\begin{figure*} 
   \centering
  \includegraphics[width=0.45\textwidth]{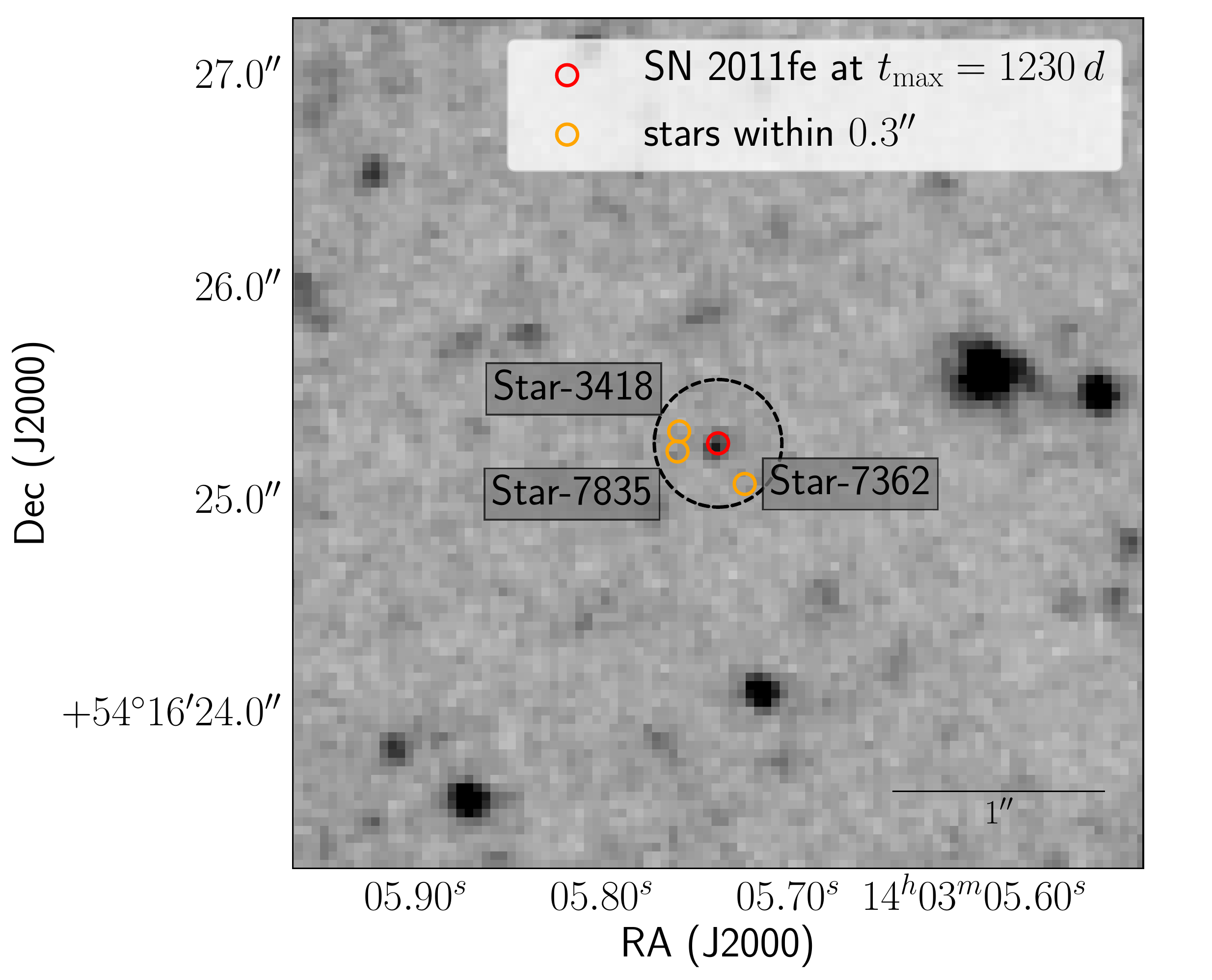} 
  \includegraphics[width=0.45\textwidth]{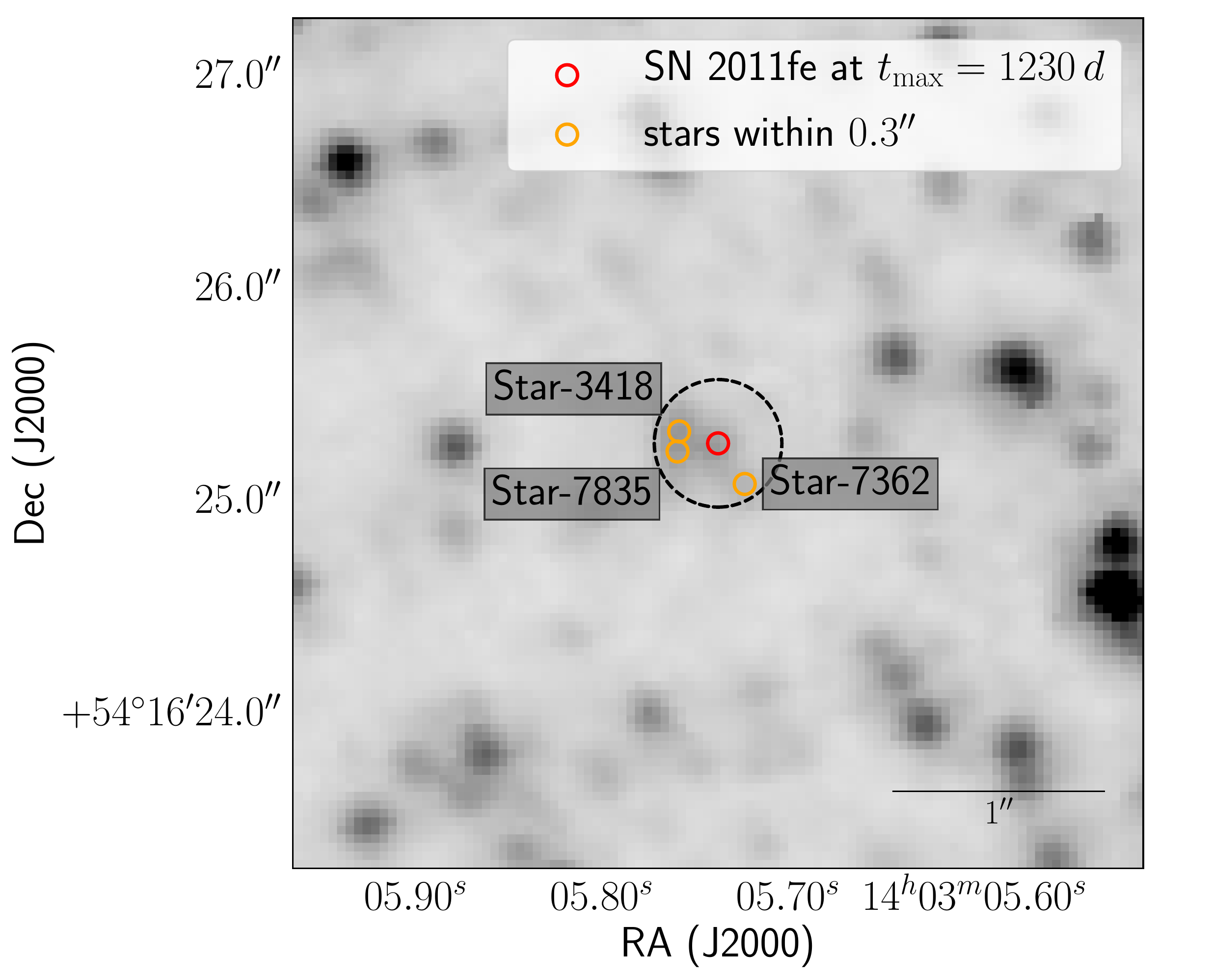} 
  \caption{Left: HST ACS F475W image of SN 2011fe at 1230\,d past maximum. Right: HST WFC3 F160W image at 1230\,d past maximum. Sources within a circle of 0.3\,arcsec radius around SN~2011fe are marked.}
 \label{fig:overview}
\end{figure*}

As reported in \citet{2015ATel.7392....1S} crowding is an issue for this source in the infrared bands (see Figure~\ref{fig:overview}). However fitting photometry for \sn{2011}{fe} on all frames simultaneously allows us to extract stellar photometry also for the under-resolved regions in the infrared \gls{hst.wfc3} images. We have checked the magnitudes of nearby stars to ensure that the flux of these remains constant over time and thus do not contribute to the supernova measurement. 

We verified the \gls{dolphot} uncertainties by checking the scatter of flux measurements of 1073 stars around the supernova. The results show that \gls{dolphot} often overestimated the uncertainty by a factor of 1.5 (in roughly 2/3 of the cases) or are close to the uncertainty. This suggests that the uncertainty determination is reliable. 

For two of our measurements the uncertainties are of similar magnitude as the measurements themselves (indicated in the table). We used three times this uncertainty and treated this as an upper limit. In all future processing, the bounded errors are treated as being normally distributed while the upper limits are treated as having uniform probability distributions between [0, upper limit]. 

The final photometry for \sn{2011}{fe} is shown in Table~\ref{tab:11fe_phot} and the synthetic evolving SED is shown in Figure~\ref{fig:synth_spec_phot}.

\subsection{Gemini Photometry}

We obtained multicolour photometry using \gls{gmosn} mounted on the Gemini North telescope. Images were taken in the $g$, $r$, $i$, and $z$ bands under the program GN-2014A-Q-24.  The first epoch (nights of 2014 March 7, 27, and 28) has already been published by \citet{2014ApJ...796L..26K}, while the second epoch (night of 2014 June 27) is presented here. 

The data were pre-reduced with the \textsc{geminiutil}\footnote{{\url{http://github.com/geminiutil/geminiutil}}} package following standard procedures. After careful inspection of the \glsdesc{wcs}, the images were aligned and combined using \gls{swarp}. In a final step, we adjusted the astrometric calibration to match \gls{hst.acs} observations. We undertook the same operations on SDSS-calibrated stars \citep{2011ApJS..193...29A} in the field and then used those for the calibration. 

Subsequently, we performed our measurements using \gls{psf} photometry with the \textsc{snoopy} package. This package is a compilation of \gls{iraf} tasks optimized for \gls{sn} photometry, developed by F.\,Patat and E.\,Cappellaro.
\textsc{snoopy} constructs
the \gls{psf} by selecting several clean unblended stars and then performs \gls{psf} photometry on the \gls{sn} itself. The instrumental SN magnitudes were finally calibrated to the Sloan photometric system \citep{1996AJ....111.1748F} using tabulated atmospheric extinction coefficients and the nightly zero points derived from our standard-field observations. 

To get a better estimate of the uncertainty of the photometric measurement, \textsc{snoopy} uses an artificial star experiment.

\begin{table*}
\caption{Photometry for \sn{2011}{fe}. }
\label{tab:11fe_phot}

\begin{tabular}{rrllrccccc}
\hline
 Visit ID &  $t_\textrm{max}$ & Instrument &  Filter &  $t_\textrm{exp}$ &  exposures &          flux & $\sigma_\textrm{flux}$ &        mag & $\lambda_\textrm{pivot}$ \\
 --&$[\mathrm{d}]$&--&--&$[\mathrm{s}]$&--&$[\mathrm{{erg}\,{\AA^{-1}\,s^{-1}\,cm^{-2}}}]$&$[\mathrm{{erg}\,{\AA^{-1}\,s^{-1}\,cm^{-2}}}]$&Vega&$[\mathrm{\AA}]$\\
\hline
       -2 &               929 &     GMOS-N &       $g$ &               900 &          5 & \num{2.6e-18} &          \num{7.5e-19} & \num{23.20} &               \num{4729} \\
       -2 &               929 &     GMOS-N &       $r$ &               900 &          5 & \num{5.2e-19} &          \num{7.1e-20} & \num{24.31} &               \num{6291} \\
       -2 &               929 &     GMOS-N &       $i$ &               900 &          5 & \num{3.4e-19} &          \num{6.1e-20} & \num{24.32} &               \num{7774} \\
       -2 &               929 &     GMOS-N &       $z$ &              1800 &         10 & \num{1.9e-19} &          \num{3.2e-20} & \num{24.48} &               \num{9695} \\
       -1 &              1021 &     GMOS-N &       $g$ &               900 &          5 & \num{1.4e-18} &          \num{1.7e-19} & \num{23.84} &               \num{4729} \\
       -1 &              1021 &     GMOS-N &       $r$ &              1800 &         10 & \num{3.2e-19} &          \num{7.6e-20} & \num{24.82} &               \num{6291} \\
       -1 &              1021 &     GMOS-N &       $i$ &              1800 &         10 &   \num{3.0e-19} &          \num{6.4e-20} & \num{24.44} &               \num{7774} \\
       -1 &              1021 &     GMOS-N &       $z$ &              3600 &         20 & \num{1.9e-19} &          \num{1.6e-20} & \num{24.46} &               \num{9695} \\
        1 &              1125 &       WFC3 &   F438W &              1998 &          3 & \num{7.6e-19} &          \num{4.5e-20} & \num{24.85} &               \num{4326} \\
        1 &              1125 &       WFC3 &   F555W &              1848 &          3 & \num{6.4e-19} &          \num{2.2e-20} & \num{24.47} &               \num{5309} \\
        1 &              1125 &       WFC3 &  F600LP &              4200 &          6 & \num{1.8e-19} &          \num{8.7e-21} & \num{24.75} &               \num{7506} \\
        1 &              1125 &       WFC3 &   F110W &              1598 &          2 & \num{8.5e-20} &            \num{4.0e-21} & \num{24.20} &              \num{11534} \\
        1 &              1125 &       WFC3 &   F160W &              1398 &          2 & \num{1.7e-19} &          \num{4.6e-21} & \num{22.36} &              \num{15369} \\
        2 &              1232 &       WFC3 &   F336W &              2850 &          3 & \num{4.8e-20} &          \num{2.9e-20} & \num{27.06} &               \num{3354} \\
        2 &              1232 &        ACS &   F475W &               360 &          3 &   \num{4.0e-19} &          \num{4.7e-20} & \num{25.30} &               \num{4744} \\
        2 &              1232 &        ACS &   F625W &               660 &          3 & \num{1.2e-19} &          \num{1.9e-20} & \num{25.72} &               \num{6310} \\
        2 &              1232 &        ACS &   F775W &               750 &          3 & \num{1.5e-19} &          \num{1.6e-20} & \num{24.83} &               \num{7693} \\
        2 &              1232 &       WFC3 &   F105W &               965 &          3 & \num{4.2e-20} &          \num{8.3e-21} & \num{25.28} &              \num{10552} \\
        2 &              1232 &       WFC3 &   F125W &               828 &          3 & \num{7.2e-20} &            \num{8.0e-21} & \num{24.07} &              \num{12486} \\
        2 &              1232 &       WFC3 &   F160W &               828 &          3 & \num{1.1e-19} &          \num{7.7e-21} & \num{22.78} &              \num{15369} \\
        3 &              1304 &       WFC3 &   F438W &              2768 &          3 & \num{4.6e-19} &          \num{3.2e-20} & \num{25.41} &               \num{4326} \\
        3 &              1304 &       WFC3 &   F555W &              3247 &          3 & \num{2.8e-19} &          \num{1.2e-20} & \num{25.36} &               \num{5309} \\
        3 &              1304 &       WFC3 &  F600LP &              5372 &          6 & \num{9.4e-20} &          \num{6.8e-21} & \num{25.44} &               \num{7506} \\
        3 &              1304 &       WFC3 &   F110W &              1598 &          2 & \num{3.9e-20} &          \num{3.4e-21} & \num{25.05} &              \num{11534} \\
        3 &              1304 &       WFC3 &   F160W &              1398 &          2 & \num{8.5e-20} &            \num{4.0e-21} & \num{23.09} &              \num{15369} \\
        4 &              1403 &       WFC3 &   F438W &              8000 &          4 & \num{3.3e-19} &          \num{1.9e-20} & \num{25.75} &               \num{4326} \\
        4 &              1403 &       WFC3 &   F555W &              3880 &          4 & \num{2.1e-19} &          \num{1.2e-20} & \num{25.67} &               \num{5309} \\
        4 &              1403 &       WFC3 &  F600LP &              5905 &          4 & \num{6.8e-20} &          \num{5.3e-21} & \num{25.80} &               \num{7506} \\
        4 &              1403 &       WFC3 &   F110W &              1598 &          2 & \num{3.7e-20} &          \num{3.4e-21} & \num{25.10} &              \num{11534} \\
        4 &              1403 &       WFC3 &   F160W &              1398 &          2 & \num{6.9e-20} &          \num{3.9e-21} & \num{23.30} &              \num{15369} \\
        5 &              1461 &        ACS &   F475W &               800 &          4 &   \num{2.0e-19} &          \num{3.1e-20} & \num{26.03} &               \num{4744} \\
        5 &              1461 &        ACS &   F625W &              1600 &          4 & $\leq$\num{3.7e-20} & & $\geq$\num{26.99} &               \num{6310} \\
        5 &              1461 &        ACS &   F775W &              2202 &          4 & \num{8.7e-20} &          \num{9.3e-21} & \num{25.41} &               \num{7693} \\
        5 &              1461 &       WFC3 &   F105W &              1931 &          6 & $\leq$\num{2.8e-20}& & $\geq$\num{25.70} &              \num{10552} \\
        5 &              1461 &       WFC3 &   F125W &              1587 &          6 & \num{4.2e-20} &            \num{8.0e-21} & \num{24.66} &              \num{12486} \\
        5 &              1461 &       WFC3 &   F160W &              1655 &          6 & \num{5.6e-20} &          \num{7.3e-21} & \num{23.53} &              \num{15369} \\
        6 &              1623 &       WFC3 &   F438W &              2779 &          3 & \num{1.7e-19} &          \num{2.4e-20} & \num{26.46} &               \num{4326} \\
        6 &              1623 &       WFC3 &   F555W &              1362 &          3 & \num{9.9e-20} &          \num{1.4e-20} & \num{26.49} &               \num{5309} \\
        6 &              1623 &       WFC3 &  F600LP &              4280 &          3 & \num{4.3e-20} &          \num{4.3e-21} & \num{26.30} &               \num{7506} \\
        6 &              1623 &       WFC3 &   F110W &              1598 &          2 & \num{2.3e-20} &          \num{2.8e-21} & \num{25.63} &              \num{11534} \\
        6 &              1623 &       WFC3 &   F160W &              1398 &          2 & \num{3.9e-20} &          \num{3.6e-21} & \num{23.92} &              \num{15369} \\
\end{tabular}
\end{table*}

\section{Analysis}
\label{sec:analysis}
Our analysis consists of three distinct actions. First we combine the individual photometric measurements, reconstructing a quasi-bolometric flux that is comparable cross-epoch (despite the use of different filter sets - see Figure~\ref{fig:synth_spec_filters}). Then we construct a light curve model with various options and finally we explore the parameter space of the model given our observed data.

\subsection{Reconstructing a quasi-bolometric flux}

For our analysis we compare the complete flux emerging in the UVOIR bands with the energy produced by radioactive decay. Thus we aimed to create a measure of this UVOIR flux that is consistent across all epochs (and taking into account filter overlap). To do this we scale a synthetic spectrum to our observed photometry and use the integrated flux of this spectrum for our comparison \citep[using the nomenclature and technique of][aptly named quasi-bolometric flux]{2017ApJ...841...48S}. We only have one measurement in the $U$ band (F336W) and thus we exclude this band from the reconstruction as our main goal is to explore the flux evolution.

\begin{figure*} 
   \centering
   \includegraphics[width=1\textwidth]{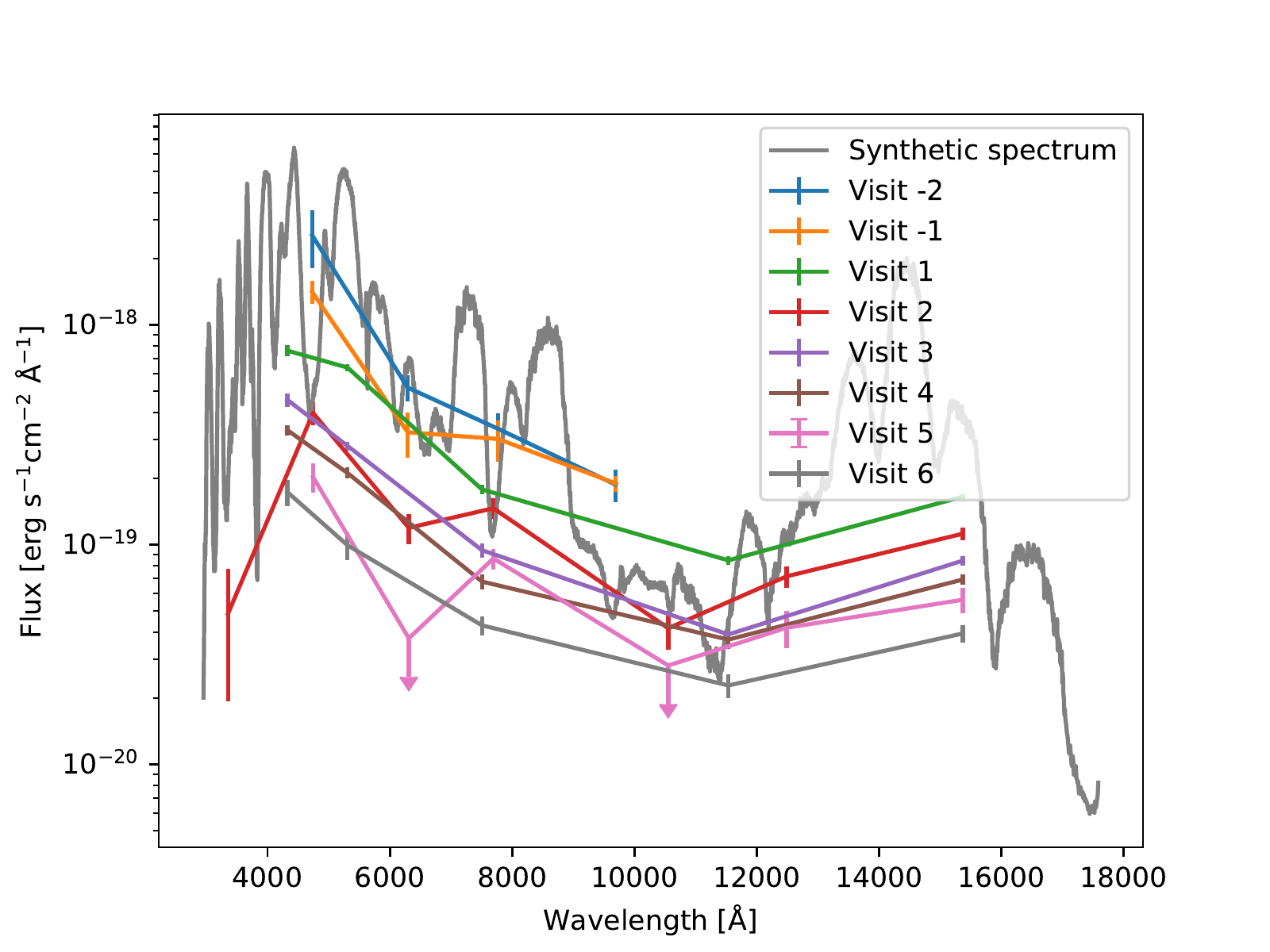} 
   \caption{The lines show the measured Flux in the different bands. For a better understanding of the SED we have also added the synthetic spectrum in arbitrary flux scaling \citep{2015ApJ...814L...2F}.}
   \label{fig:synth_spec_phot}
\end{figure*}

\begin{figure} 
   \centering
   \includegraphics[width=0.49\textwidth]{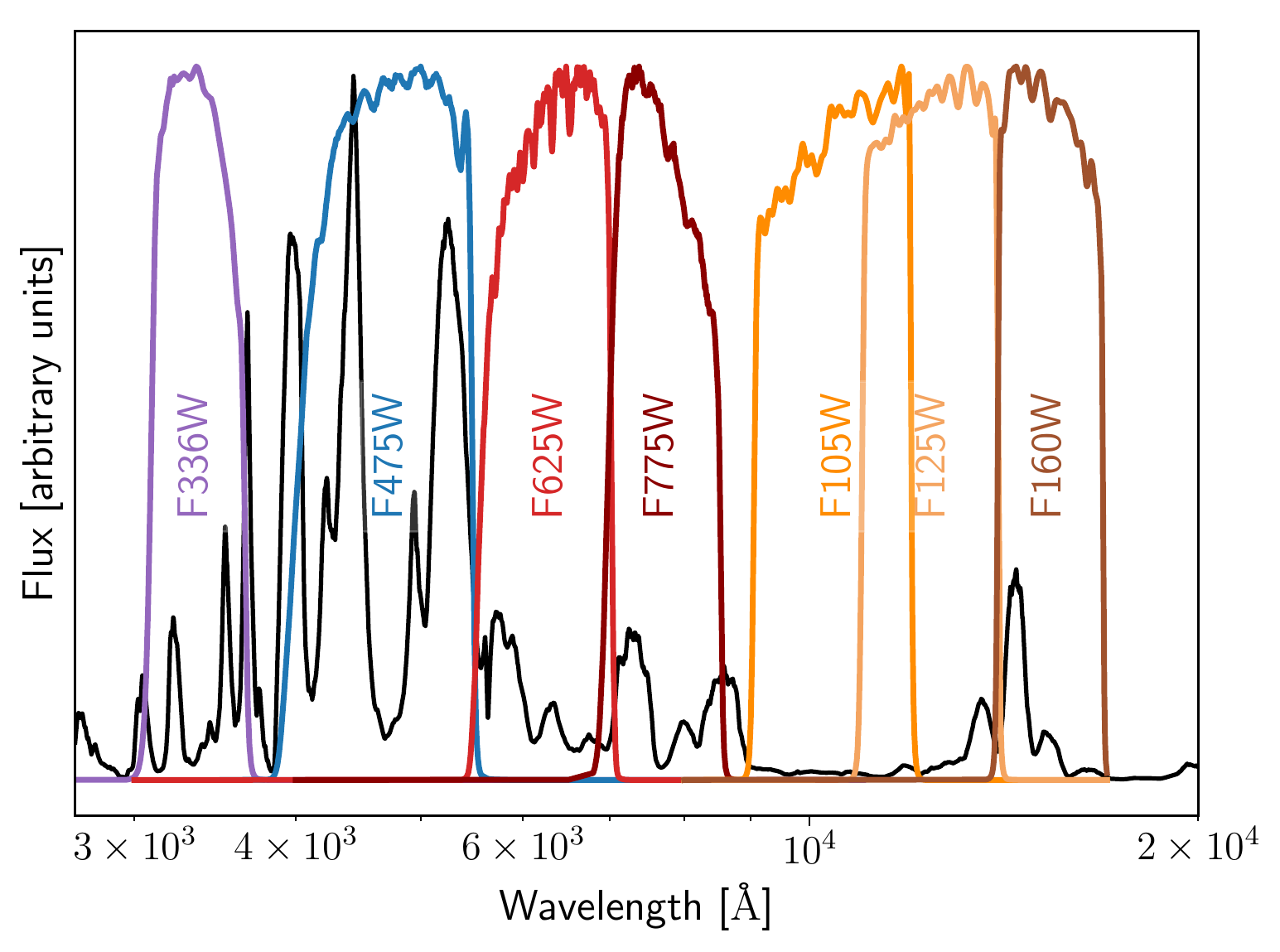} 
   \includegraphics[width=0.49\textwidth]{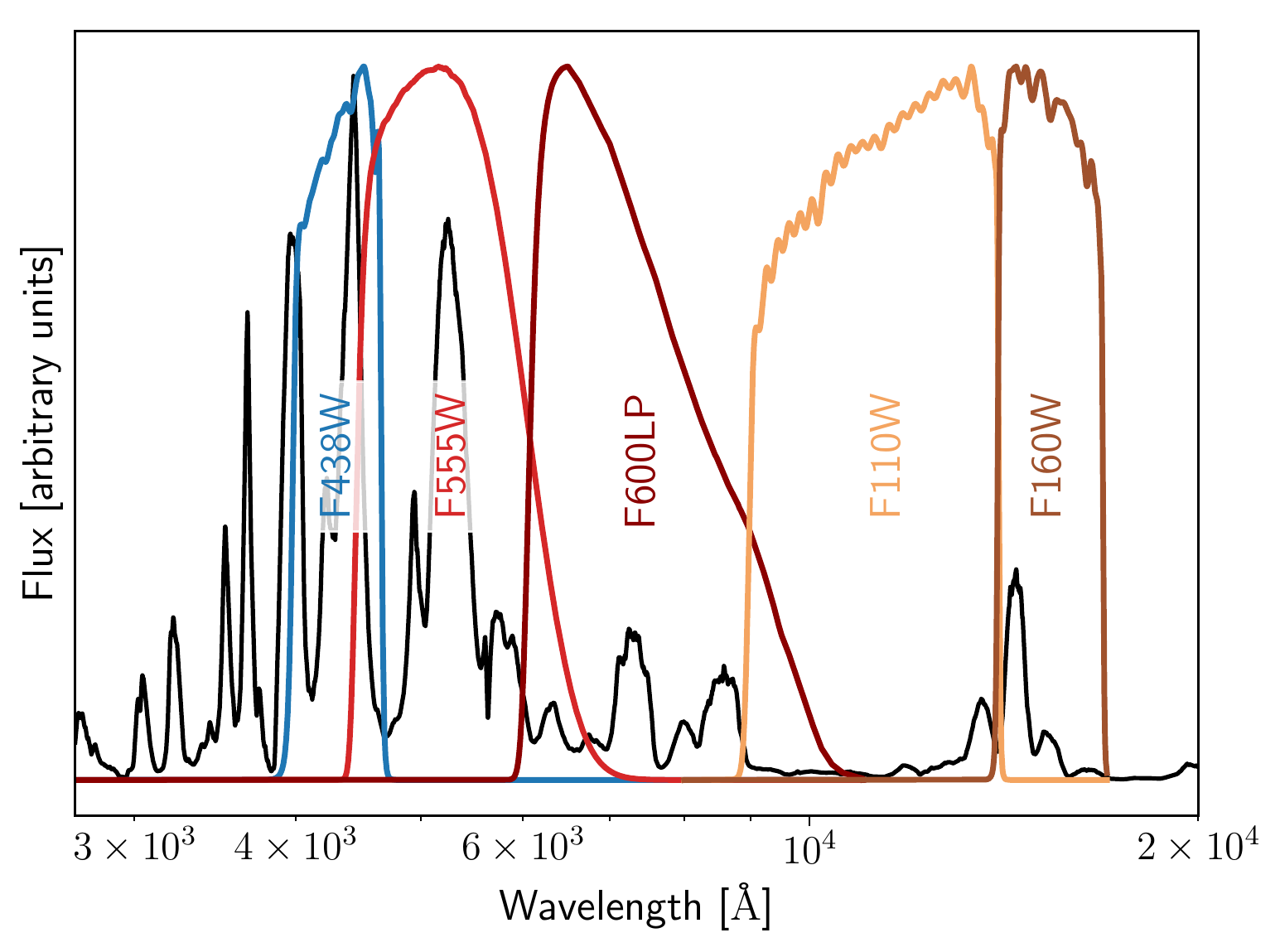} 
   \caption{The figures show the synthetic spectrum by \citet{2015ApJ...814L...2F} with the filter sets used by this team (top figure) and \citet{2017ApJ...841...48S} (bottom figure) superimposed.}
   \label{fig:synth_spec_filters}
\end{figure}

\citet{2015ApJ...814L...2F} have modeled the physical processes and published a synthetic spectrum for \sn{2011}{fe} around 1000 days past explosion that is ideal for our experiment. Figure~\ref{fig:synth_spec_filters} shows the synthetic spectrum as well as the filter sets used. We perform synthetic photometry on this synthetic spectrum using the filter curves provided by STScI\footnote{
for WFC3 \url{http://www.stsci.edu/hst/wfc3/ins_performance/throughputs/Throughput_Tables}\\
for ACS \url{http://www.stsci.edu//hst/acs/analysis/throughputs/tables}}.  An initial comparison of the synthetic spectrum with the observed flux suggests that the spectrum is a relatively good approximation for the true emission (see Figure~\ref{fig:synth_spec_phot}). However, in the blue part of the spectrum the synthetic photometry starts to deviate.

To alleviate this deviation, we adopt the method by \citet{2017MNRAS.468.3798D} and warp the synthetic spectrum using linear interpolation between the different bands. For this purpose, we compute synthetic photometry of the synthetic spectrum and divide this with our observed photometry. We then linearly interpolate between these scalings and multiply the spectrum with this warp. This procedure is repeated until the maximum relative difference is less than 0.0001 in flux (among all bands). We estimate our uncertainties by repeating this experiment by drawing 1000 observed data points (randomly sampled using their uncertainties) and then calculate the quasi-bolometric flux measurement by integrating them. Our final integral of the warped synthetic spectrum is from 4000\,\AA\ to 16000\,\AA\ (thus excluding F336W). 
We believe the given process to be the optimal way to sum the flux given the unknown flux distribution and the different filter sets used at different epochs (we will discuss some implications in Section~\ref{sec:discussion}). 
The final quasi-bolometric luminosities are shown in Table~\ref{tab:quasi_bolometric}.

\begin{table}
\caption{The reconstruction of the quasi-bolometric light curve}
\label{tab:quasi_bolometric}

\begin{tabular}{rrrr}
\hline
 Visit ID & $t_\textrm{max}$ &  flux density & $\sigma_\textrm{flux density}$ \\
 -&[$d$]&[$\mathrm{{erg}\,{s^{-1}\,cm^{-2}}}$]&[$\mathrm{{erg}\,{s^{-1}\,cm^{-2}}}$]\\
\hline
        -2 &              929 & \num{8.4e-15} &                  \num{1.4e-15} \\
        -1 &             1021 & \num{6.1e-15} &                  \num{4.6e-16} \\
        1 &             1125 & \num{2.8e-15} &                  \num{5.7e-17} \\
        2 &             1232 & \num{1.7e-15} &                  \num{9.6e-17} \\
        3 &             1304 & \num{1.4e-15} &                    \num{4.0e-17} \\
        4 &             1403 & \num{1.1e-15} &                  \num{3.3e-17} \\
        5 &             1461 & \num{8.4e-16} &                  \num{7.1e-17} \\
        6 &             1623 &   \num{6.0e-16} &                  \num{3.5e-17} \\
\end{tabular}
\end{table}

\subsection{Light-curve Model \& Likelihood}

\begin{figure} 
   \centering
   \includegraphics[width=0.5\textwidth]{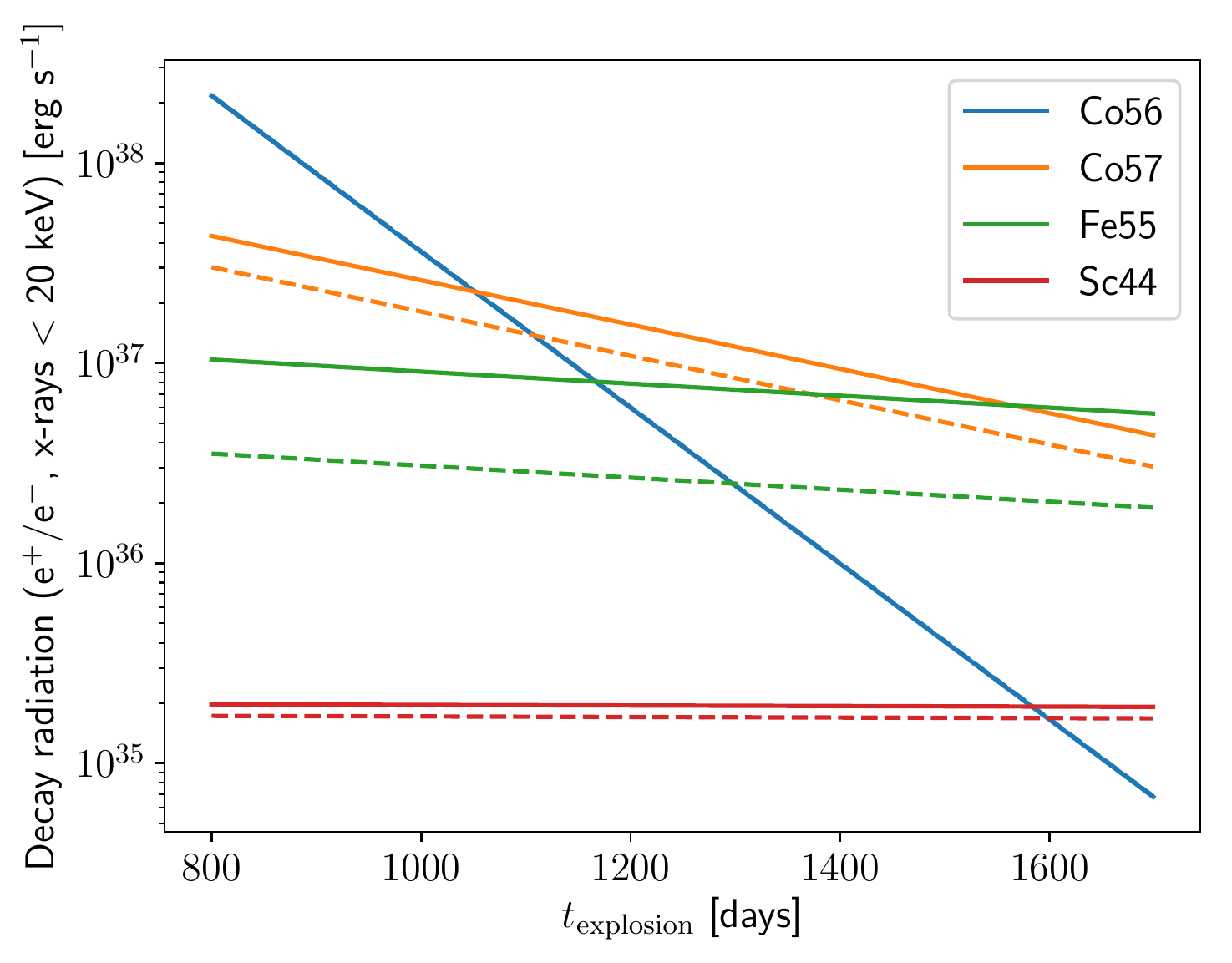} 
   \caption{Decay radiation of four isotopes producing the highest energies when considering electrons/positrons and x-rays up to 20\,keV. The solid lines are the N100 model. The dashed lines are the merger model.}
   \label{fig:decay_energy}
\end{figure}

For the model of our quasi-bolometric light curve we start by considering only radioactive decay as an energy source and assume that the energy is deposited instantaneously. Figure~\ref{fig:decay_energy} shows the decay radiation produced by the different isotopes at late times using the abundances given in \citet{2012ApJ...750L..19R}.

The energy input for our model is based on three decay chains

\begin{itemize}
\item $\nucl{Ni}{56} \xrightarrow{t_{1/2} = 6\,\textrm{d}} {\nucl{Co}{56}} \xrightarrow{t_{1/2} = 77\,\textrm{d}} {\nucl{Fe}{56}}$
\item $\nucl{Ni}{57} \xrightarrow{t_{1/2} = 36\,\textrm{h}} {\nucl{Co}{57}} \xrightarrow{t_{1/2} = 272\,\textrm{d}} {\nucl{Fe}{57}}$
\item $\nucl{Co}{55} \xrightarrow{t_{1/2} = 18\,\textrm{h}} {\nucl{Fe}{55}} \xrightarrow{t_{1/2} = 985\,\textrm{d}} {\nucl{Mn}{55}}$
\end{itemize}

which produce the bulk of the energy at the epochs under consideration. For given input masses of these isotopes, we calculate the current masses using  the \gls{pyne} software package. We then calculate the decay radiation energy \citep[data from][]{1998NDS....85..415B,2008NDS...109..787J,2011NDS...112.1513J} taking into account all electrons/positrons (beta-decay, Auger and internal-conversion electrons) as well as all electromagnetic radiation up to 20 keV. We then parameterize the fraction of the decay energy that is radiated outside the observed bands and assume the rest of it is radiated within our observed window of 4000\,--\,16000\,\AA. Finally we scale the energy using the distance to the supernova. In total this gives five parameters (\Ni, \Ni[57] and \Co[55] masses, the fraction outside the observed bands, and the distance). 


\subsection{Exploring the parameter space}
\label{sec:expl_param}
We employ a Bayesian framework to explore the parameter space. Thus, we assume a Gaussian prior with $(0.5\pm0.1)\,\msun$ for \Ni[56] from \citet[][]{2015MNRAS.454.3816C} with a more conservative uncertainty, uniform priors of $0 - 0.1$\,\msun\ for \Ni[57] and \Co[55],  a Gaussian prior centered on 0.73 for the fraction of energy radiated outside of the observed bands falling off with a sigma of 0.1 towards a higher fraction and with a sigma of 0.3 towards a lower fraction \citep[given a model by][and capped between 0 and 1]{2015ApJ...814L...2F} and a prior of ($6.4\pm0.6$)\,Mpc for the distance \citep{2011ApJ...733..124S}. Specifically, for the fraction outside of the observed bands, we assume time independence which is likely a good assumption as the plasma state will only change extremely slowly at these very late times. Our likelihood is a simple $\chi^2$-likelihood that compares the generated models to the data. The parameter space is sampled using the \gls{multinest} algorithm and using the implementation available at \url{https://github.com/kbarbary/nestle}).

\begin{figure*} 
   \centering
   \includegraphics[width=0.98\textwidth]{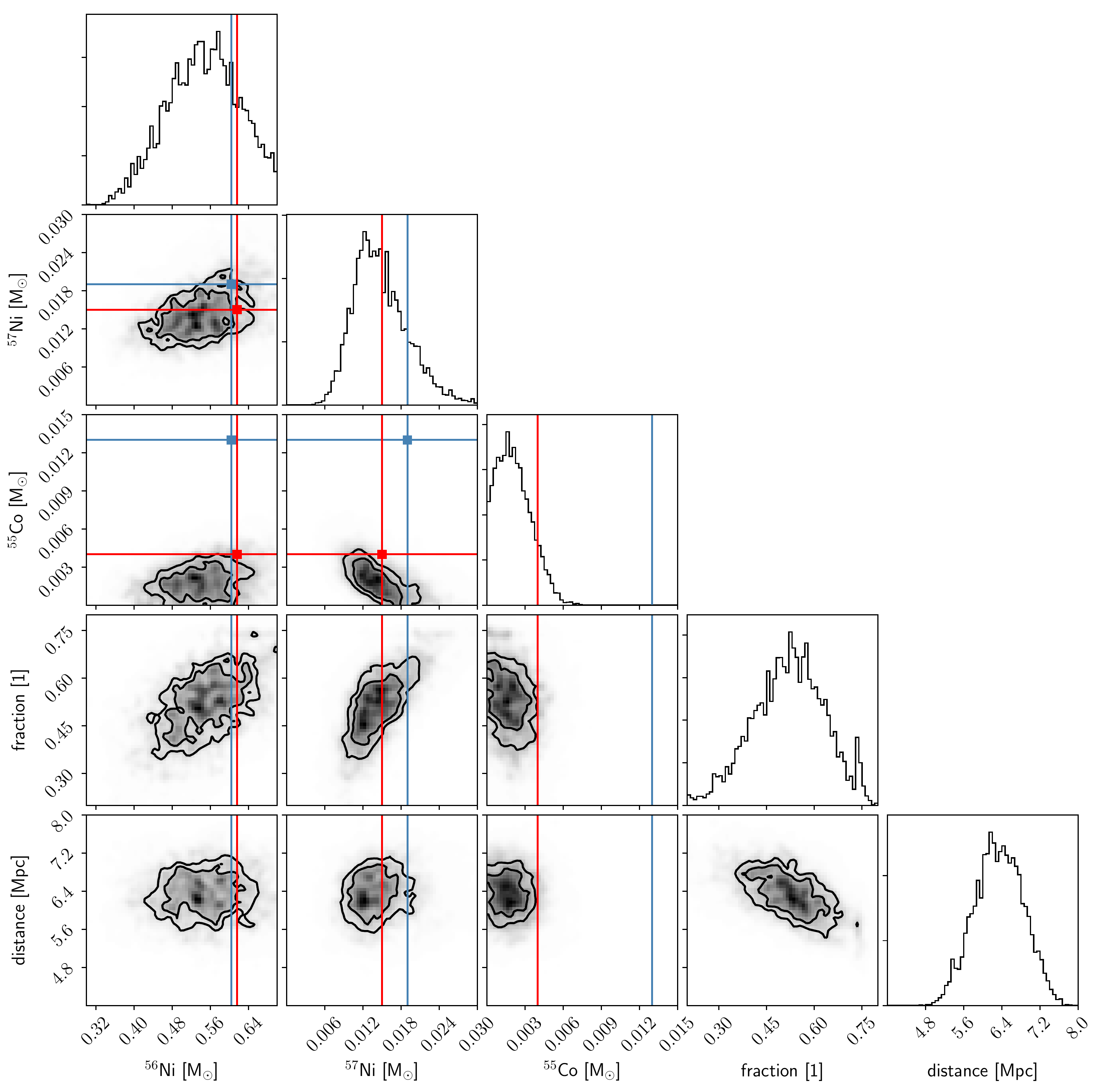} 
   \caption{Corner plot of our model parameters with 68\% and 95\% confidence intervals  \citep[using \textsc{corner.py};][]{dan_foreman_mackey_2016_53155}. The contours rely on a smoothed probability density function for better visualization. The values from \citet{2012ApJ...750L..19R} for the merger model have been marked in red and for the N100 model in blue (with a \Ni[56] mass which is slightly on the high side for SN~2011fe).}
   \label{fig:corner_plot}
\end{figure*}

\begin{figure} 
   \centering
   \includegraphics[width=0.5\textwidth]{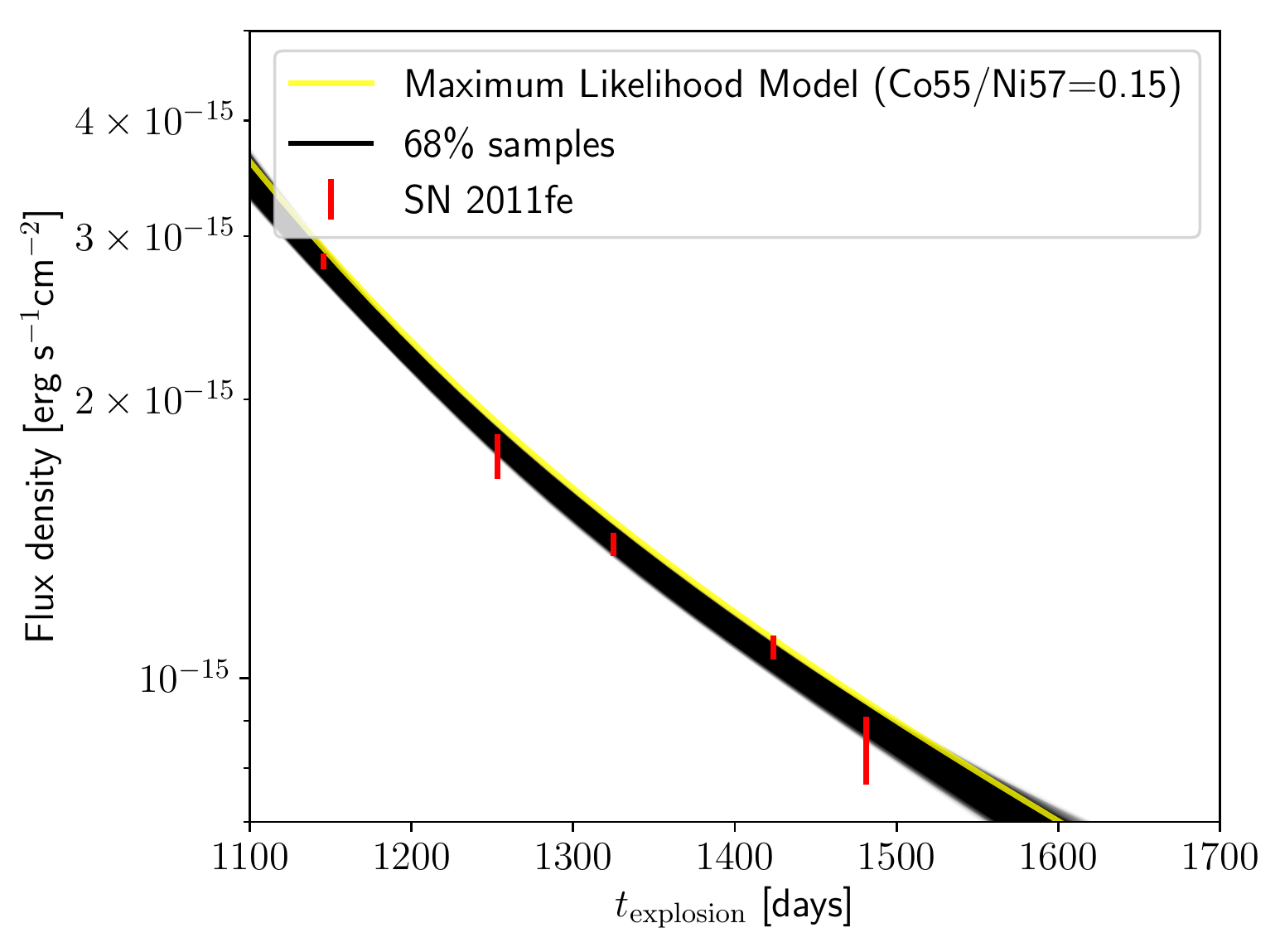} 
   \caption{A fit of the radioactive model to the data. We show the maximum likelihood realization as well as all curves within the 68\%-quantile.}
   \label{fig:simple_model}
\end{figure}

We present the posterior probability in  Figure~\ref{fig:corner_plot} including two confidence intervals (68\%, 95\%), while in Figure~\ref{fig:simple_model} we present all model light curves lying within the 68\% confidence region.

Furthermore, we explored the sensitivity of our fit to some very simple prescriptions for the  electron/positron escape and the freeze-out effect.  We parametrized the freeze-out effect as an additional energy source contributing to the late-time light curve and simply assumed this effect follows the  $n_e \times n_\textrm{ion} \times V \propto t^{-3}$ relation. For the electron positron escape, we assumed a simple parametrization (escape fraction = $1 - \exp{(-(t/t_d)^2)}$) choosing $t_d=1200~d$ \citep[assuming a weak magnetic field in addition to that discussed in][]{2017arXiv170206702J}. 

For both these effects, we focused on just finding a good fit and opted for a differential evolution algorithm to fit the appropriate parameters (isotopic masses, observed fraction, the scaling of the freeze out).

\section{Discussion}
\label{sec:discussion}

As discussed above, we aim to use the measured energy output, relate this to the current energy injection rate and use this to determine isotopic ratios which in turn may be used as discriminants of explosion mass and mechanism \citep[as predicted by][]{2012ApJ...750L..19R}. However, in addition to the radioactive decay, there are several physical effects that if not taken into account might skew the isotopic abundance measurements (and thus any conclusions we draw). External sources of light contamination (such as companions, light echo, etc.) have been discussed in detail by both \mbox{\citet{2017ApJ...841...48S}} and \mbox{\citet{2017MNRAS.468.3798D}}. 

For the internal process of converting the radioactively produced energy into observable light, \citet{2017ApJ...841...48S} assume instantaneous deposition and immediate re-radiation of the radioactively produced energy. \citet{2017MNRAS.468.3798D} have also taken into account a form of parametrized escape of (electrically charged) leptons and gamma-rays (as well as hinting at potential problems due to freeze-out effects). 

The uncertain magnitude of these effects results in large systematic uncertainties. Here, we will focus our discussion on some specific sources of these systematic uncertainties: 1) creation of a quasi-bolometric light curve 2) the limited wavelength range we observe 3) energy sources/sinks other than radioactive decay. 

\paragraph*{Quasi-bolometric flux} Our measurements are energy output in different wavelength bands. First, we generate a derived quantity (quasi-bolometric flux)  that can be compared between epochs despite the different filter sets that have been used. We construct this quasi-bolometric flux using a probable SED by \mbox{\citet{2015ApJ...814L...2F}} which we warp slightly to reflect the mismatch when compared to the measured SED. We believe that this gives us a reliable estimate of a quasi-bolometric flux, but acknowledge that this is dependent on our choice of the synthetic spectrum (as well as the warping technique). Our experiments using simpler techniques (e.g. linear interpolation between the filters' effective wavelengths) show that they differ by a few percent (to a maximum of 8\%) and thus have an effect on the overall uncertainty budget but do not dominate it.

\paragraph*{Radiation outside the observed window} Another major uncertainty comes from the fact that we only capture flux in a UVOIR window. Specifically, two factors have an important influence on the precise measurement of the isotopic abundances. The amount of the flux outside of our observed window and more importantly if the fraction of this flux compared to the observed one is constant.  

Radioactive decay in \sneia produces decay electrons/positrons, X-rays, and $\gamma$-rays. For the late phases considered here the ejecta are completely transparent to $\gamma$-rays, whereas lower-energy X-rays contribute. These are absorbed and produce fast electrons. This energetic electron/positron pool is supplemented by the beta decays, internal conversion and Auger electrons. Figure 5 in \citet{2015ApJ...814L...2F} shows that a majority of this energy goes into the thermal pool ($>60\%$) while the other part goes into non-thermal ionization and excitation (being shown for a pure iron plasma). This leads to an electron temperature of roughly 100\,K for the outer layers and a few 100\,K for the core.

The thermal pool at these late times is only able to excite fine-structure lines in the mid-/far-IR. This leads to cooling via the prominent Fe and Si lines between 20\,\micron\,--\,35\,\micron\ \citep[see ][for details]{2015ApJ...814L...2F}. The presented observations are evidence that other processes must be at play to excite lines in the UVOIR bands. As discussed a non-negligible fraction of the energetic electrons/positrons do convert their energy into non-thermal excitation and ionization which is primarily radiated via UV lines. The lack of strong UV flux in epoch 2 of the observations, however, suggests that these are reprocessed \citep[as predicted by ][]{2015ApJ...814L...2F}.

\begin{table*}
\begin{tabular}{llll}
\toprule
{} & $\log_{10}~\Ni[56]/\Ni[57] $ & $ \log_{10}~\Ni[56]/\Co[55]$ & $\log_{10}~\Co[55]/\Co[57]$ \\
\midrule
N100 [theoretical]               &                          1.7 &                          1.5 &                         0.2 \\
merger [theoretical]             &                          1.6 &                          2.2 &                      $-0.6$ \\
pure radioactive (68\% quantile) &                    1.4 - 1.9 &                    2.0 - 5.0 &               $-3.1 - -0.6$ \\
e+/e- escape                     &                            3 &                         1.17 &                         1.8 \\
Shappee et al. 2017              &                $1.52 - 1.65$ &                     $> 2.35$ &                    $< -0.7$ \\
Dimitriadis et al. 2017 (Case 2) &                          1.2 &                          1.5 &                      $-0.3$ \\
\bottomrule
\end{tabular}
\caption{Comparison between the different measurements. Freeze-out was excluded as it only has \Ni[56] measuerment consistent with the observations, but zero for all other isotopes.}
\label{tab:isotope_compare}
\end{table*}

\paragraph*{A simple pure radioactivity-driven model}
We have compared this model to the theoretical models of \citet{2012ApJ...750L..19R} and the measurements of \cite{2017ApJ...841...48S, 2017MNRAS.468.3798D} in Table~\ref{tab:isotope_compare}.  Our pure radioactivity based modelling gives results that are in agreement with the values presented in \citet{2017ApJ...841...48S}. We are not consistent with \citet[][]{2017MNRAS.468.3798D} for the pure radioactivity model as they include additional physics (that we will also add in subsequent steps).

However, as discussed, it is unlikely that this simple radioactivity-driven model does capture all the physical processes that lead to the observed light curve. 
Other processes do have a significant impact on the shape of the light curve (as described in the following). Any constraints on any isotopic abundances thus carry large systematic uncertainties. 

\paragraph*{Energy sinks/sources }
\citet[][]{2015ApJ...814L...2F} suggest that the non-thermally excited ions might absorb a large fraction of the energy but due to the low density, the rate of recombination is slow and the release of the energy will occur much later than the injection (known as freeze-out). In effect the gas is not in ionisation equilibrium. This storage of energy will lead to a flattening of the light curve and will make any isotopic measurements using a purely radioactive decay driven model unreliable. 

\begin{figure} 
   \centering
   \includegraphics[width=0.5\textwidth]{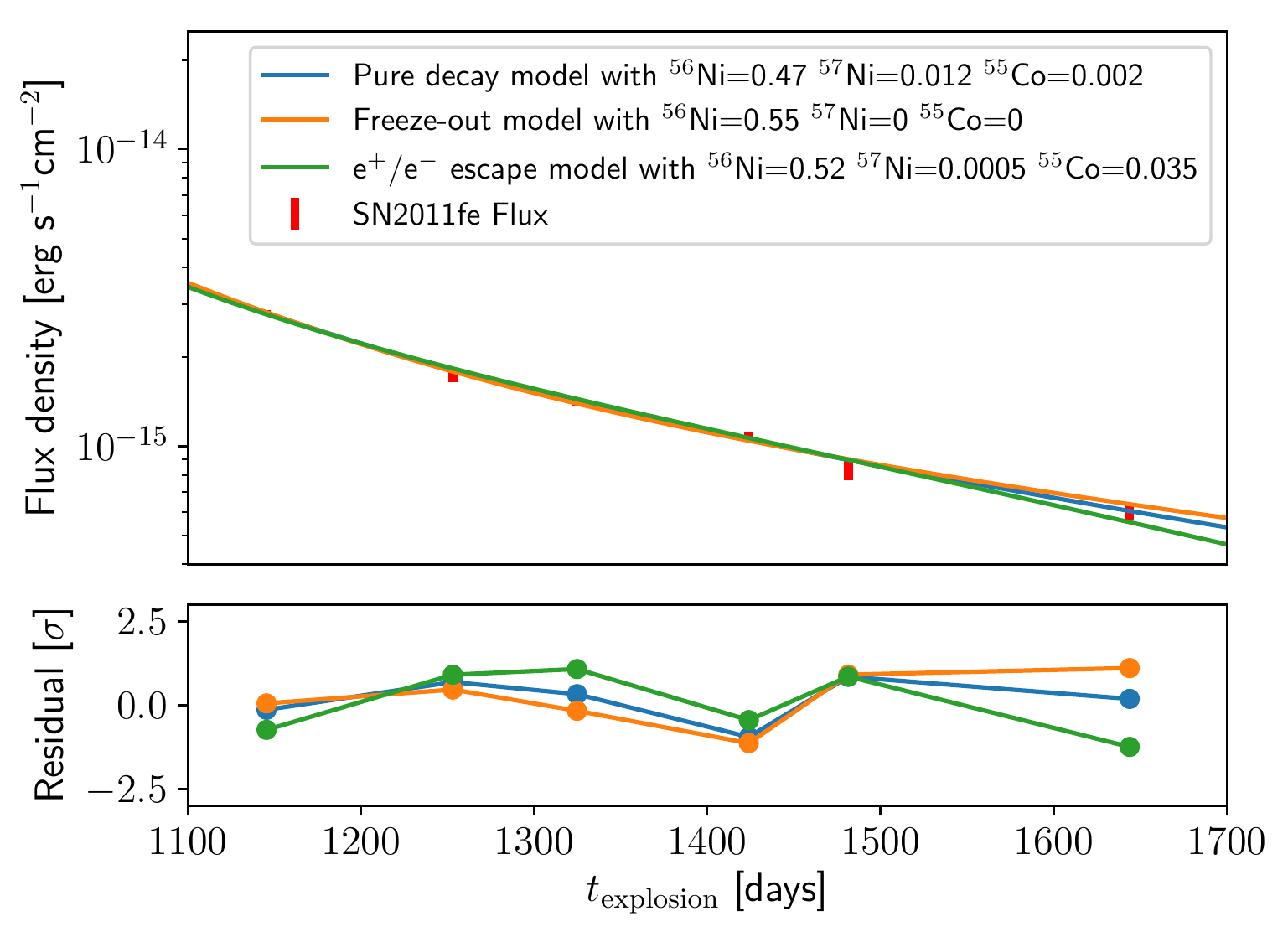} 
   \caption{We show three fits to the observed quasi-bolometric flux using three distinct models: a purely radioactive decay-driven model, a model without any \Ni[57] and \Co[55] but with an additional prescription for freeze-out, and a model that has a time-dependent electron/positron escape fraction.  All solutions are very similar (as they fit the same data) and so we plotted the fractional difference at the bottom using the radioactive model as a reference.}
   \label{fig:freezeout}
\end{figure}

We use a very simple model to test how the freeze-out effect influences the already uncertain isotopic mass determinations. Figure~\ref{fig:freezeout} shows this effect produces a good fit even with the absence of $^{57}$Ni and $^{55}$Co. This result (a mimicking of the existence of \Ni[57]) has already been discussed in \citet{1993ApJ...408L..25F} and shows how crucial a precise understanding of these late time effects is for this experiment.

\citet{2014ApJ...796L..26K} have discussed at length the influence of different magnetic field configurations on the trapping of the crucial electron/positrons \citep[also discussed in][]{2017MNRAS.468.3798D}. A complete escape of these would lead to a sudden drop in the light curve. However, a gradual loss of a fraction of these energy carriers will subtly influence the light curve and gives an additional systematic uncertainty in the isotopic abundance determination. Figure~\ref{fig:freezeout} shows a very well fitting light curve with completely flipped isotopic abundances of $^{57}$Ni (0.005) and $^{55}$Co (0.035) compared to the maximum likelihood fit of the pure radioactivity-driven model (for the functional form refer to Section~\ref{sec:expl_param}). This is consistent with Case 2 in \citet[][see Table~\ref{tab:isotope_compare}]{2017MNRAS.468.3798D}

\section{Conclusion}
\label{sec:conclusion}

In this work we present an analysis of deep HST imaging as well as Gemini imaging at late times (900 -- 1600\,days) similar to \citet{2017ApJ...841...48S} and \citet{2017MNRAS.468.3798D}. We show that one can not draw strong conclusions from the current data due to the uncertain physical processes. For example, detailed modeling of \citet{2015ApJ...814L...2F} has already shown that a simple thermal model can not explain the observed spectra \citep{2015MNRAS.448L..48T} and that the observed luminosity is likely powered by multiple effects in addition to the decay radiation. 

Using such data to determine a precise isotopic abundance does require detailed modeling of the processes that shape the observables at these late times. In particular, the freeze-out effects, a possible electron/positron escape which is linked with the magnetic field configuration \citep[see][]{1998ApJ...500..360R} and x-ray escape will be crucial to this effort. On the observational side, we believe that a successful detection of the cooling lines in the mid-IR will further strengthen the arguments brought forth by \citet{2015ApJ...814L...2F}. This can likely be achieved -- even at very late times -- with JWST. 

\section{Acknowledgements}

W.~E.~Kerzendorf was supported by an ESO Fellowship. A.~J.~Ruiter acknowledges funding through the Australian Research Council Centre of Excellence for All-sky Astrophysics (CAASTRO) through project number CE110001020. S.~Taubenberger acknowledges support by TRR33 `The Dark Universe' of the German Research Foundation (DFG).  K.~S.~Long acknowledges support through program \#13824  provided by NASA through a grant from the Space Telescope Science Institute. I.~R.~Seitenzahl acknowledges funding from the Australian Research Council Future Fellowship grant FT160100028. A. Jerkstrand acknowledges funding by the European Union as Framework Programme for Research and Innovation Horizon 2020 under Marie Sklodowska-Curie grant agreement No 702538.

Based in part on observations obtained at the Gemini Observatory, which is operated by the Association of Universities for Research in Astronomy, Inc., under a cooperative agreement with the NSF on behalf of the Gemini partnership: the National Science Foundation (United States), the National Research Council (Canada), CONICYT (Chile), Ministerio de Ciencia, Tecnolog\'{i}a e Innovaci\'{o}n Productiva (Argentina), and Minist\'{e}rio da Ci\^{e}ncia, Tecnologia e Inova\c{c}\~{a}o (Brazil). Based also on observations made with the NASA/ESA Hubble Space Telescope, obtained at the Space Telescope Science Institute, which is operated by the Association of Universities for Research in Astronomy, Inc., under NASA contract NAS 5-26555. These observations are associated with programs \#13824, \#13737, and \#14166.”

In addition to the SW packages mentioned in the paper, we used the following software \glsfirst{astropy}, \glsfirst{numpy} \glsfirst{scipy}, \glsfirst{pandas}, \gls{matplotlib} and \glsfirst{aplpy} to analyze and visualize the data. 

Finally, we would like to thank the anonymous referee and Ben Shappee for suggestions that did improve the paper.

\bibliographystyle{mnras}
\bibliography{sn2011fe_phot2}

\end{document}